\documentclass[11pt]{article}
\usepackage[latin9]{inputenc}
\usepackage{amsmath}
\usepackage{amssymb}
\usepackage{graphicx}
\usepackage{esint}
\usepackage[unicode=true,
 bookmarks=false,
 breaklinks=false,pdfborder={0 0 1},backref=section,colorlinks=false]
 {hyperref}
\usepackage{breakurl}

\makeatletter

\DeclareFontEncoding{LGR}{}{}
\DeclareRobustCommand{\greektext}{%
  \fontencoding{LGR}\selectfont\def\encodingdefault{LGR}}
\DeclareRobustCommand{\textgreek}[1]{\leavevmode{\greektext #1}}
\ProvideTextCommand{\~}{LGR}[1]{\char126#1}

\providecommand{\tabularnewline}{\\}

\@ifundefined{date}{}{\date{}}

\usepackage{esint}
\usepackage{latexsym}
\usepackage{graphicx}\usepackage{bm}\usepackage{longtable}

\usepackage{xcolor}

\@addtoreset{equation}{section}

\makeatother

\begin{document}

\title{The interaction of glueball and heavy-light flavoured meson in holographic
QCD}
\maketitle
\begin{center}
\footnote{Email: siwenli@dlmu.edu.cn}Si-wen Li\emph{$^{\dagger}$}
\par\end{center}

\begin{center}
\emph{$^{\dagger}$Department of Physics,}\\
\emph{ Dalian Maritime University, }\\
\emph{Dalian 116026, China}\\
\par\end{center}

\vspace{10mm}
\begin{abstract}
We construct the D4/D8 brane configuration in the Witten-Sakai-Sugimoto
model by introducing a pair of heavy flavour brane with a heavy-light
open string. The multiplets created by the heavy-light string acquire
mass due to the finite separation of the heavy and light flavour branes
thus they could be identified as the heavy-light meson fields in this
model. On the other hand the glueball field is identified as the gravitational
fluctuations carried by the close string in the bulk, so this model
is able to describe the interaction of glueball and heavy-light meson
through the open-close string interaction in gauge-gravity duality.
We explicitly derive the effective action for the various glueballs
and heavy-light mesons then numerically evaluate the associated coupling
constants. Afterwards the decay widths of various glueballs to the
lowest heavy-light meson, which is identified as $D^{0}$ meson, are
calculated by using our effective action. This work extends the previous
investigations of glueball in holographic QCD and it is also a further
prediction of glueball-meson interaction.
\end{abstract}
\newpage{}

\tableofcontents{}

\section{Introduction}

In nuclear physics, Quantum Chromodynamics (QCD) as the underlying
theory of the strong interaction allows to from bound states of pure
gauge bosons due to its nonabelian nature. These bound states are
named as ``glueball''s \cite{key-1,key-2,key-3} which are usually
denoted by a range of quantum numbers $J^{PC}$ where $J,P,C$ represents
total spin, parity and charge conjugation respectively. For instance,
the lowest glueball takes the quantum numbers of the vacuum as $J^{PC}=0^{++}$.
In Yang-Mills theory, glueball is believed as the only possible composite
particle states and its spectrum has been detailedly studied by using
lattice method \cite{key-4,key-5,key-6}. While the evident existence
of the glueball is expected, their experimental identification in
the hadron spectra is challenging. This difficulty is mostly due to
the mixing of glueball with $q\bar{q}$ (quark and anti-quark) states
which have the same quantum numbers of the glueball in the presence
of quarks. Nonetheless the mass of the lightest scalar glueball is
predicted to be around 1600-1800MeV by the simulations of lattice
QCD \cite{key-4,key-7}. However the coupling and decay products/widths
of glueball have not been explicitly provided by the current status
of lattice QCD since lattice theory including dynamical quarks expressed
with real-time quantities would become extremely complexed.

Additional to the lattice gauge theory, an alternatively different
approach to investigate strongly coupled quantum field theory has
been proposed in \cite{key-8,key-9} which is the famous anti-de Sitter/conformal
field theory (AdS/CFT) correspondence, or more generally, the gauge-gravity
(string) duality. In the AdS/CFT correspondence, the correlation functions
of gauge invariant operators with the limitation of large $N_{c}$
(colour number) and large \textquoteright t Hooft coupling $\lambda$
are mapped to the perturbations of the geometrical backgrounds in
classical supergravity. So the correlation functions of strongly coupled
quantum field theory could be calculated by the weakly coupled gravity
theory in which the perturbation method is valid \cite{key-10}. Significantly,
a top-down construction for QCD based on $N_{c}$ D4-branes compactified
on a circle in type-IIA string theory was proposed by Witten in 1998
\cite{key-11} in which both supersymmetry and conformal symmetry
are broken below a Kaluza-Klein mass scale $M_{KK}$, so that the
dual field theory is four-dimensional pure Yang-Mills theory in the
large-$N_{c}$ limit. By embedding $N_{f}$ pairs of coincident D8-
and anti D8-branes ($\mathrm{D}8/\overline{\mathrm{D}8}$-brane) as
flavours into this D4-brane configuration, quarks in the fundamental
representation of flavour and colour have been introduced into this
model by Sakai and Sugimoto in 2004 \cite{key-12,key-13}. Hence this
model, named as the Witten-Sakai-Sugimoto (WSS) model, becomes remarkably
successful since it almost contains all necessary ingredients of QCD
and it is able to reproduce various fundamental features of low-energy
nuclear physics with few parameters \cite{key-14,key-15,key-16,key-17,key-18,key-19,key-20,key-21}.

Particularly the glueball field in this model is identified as the
gravitational perturbations in the bulk supergravity while meson and
baryon states are created by the open strings on the flavoured $\mathrm{D}8/\overline{\mathrm{D}8}$-brane
and ``baryonic D4-branes''\footnote{We use ``baryonic D4-brane'' to refer to the D4-branes wrapped on
$S^{4}$ which create the baryon states in this model. The details
could be reviewed in \cite{key-20,key-22}}. So this model naturally includes the interaction of glueball and
hadron which is nothing but the open-close string interaction in the
viewpoint of string theory. Specifically the mass spectrum of glueball
could be obtained by evaluating the eigen equation of the bulk graviton
and the glueball-hadron interaction would arise in the Dirac-Born-Infield
(DBI) action of the D-brane once the gravitational perturbation is
involved \cite{key-23,key-24,key-25,key-26,key-27}. 

In this paper, we will focus on the glueball-meson interaction by
including the heavy flavour in order to extend the investigation in
\cite{key-25,key-26} with this model. Since the fundamental quarks
in this model are created by the open string with one end attached
to the D4-brane and the other end to the $\mathrm{D}8/\overline{\mathrm{D}8}$-brane
which has zero length, it implies the fundamental quarks are massless.
Thus the meson states consisted of these massless quarks are usually
identified as the light flavour mesons, such as $\pi,\rho$ meson.
To involve the heavy flavour, we are going to employ the mechanism
proposed in \cite{key-28,key-29}, that is to introduce another pair
of flavoured $\mathrm{D}8/\overline{\mathrm{D}8}$-brane (as heavy
flavour branes) separated from the other coincident $\mathrm{D}8/\overline{\mathrm{D}8}$-brane
(as light flavour branes) with a open string (as heavy-light string)
stretched between them as illustrated in Figure \ref{fig:2}. Accordingly
the multiplets created by the heavy-light string acquire mass due
to the finite separation of the heavy and light flavour branes. So
the model is able to include the heavy-light flavoured mesons with
massive quarks in this approach and the mechanism is recognized as
the ``Higgs mechanism'' in string theory \cite{key-30,key-31}.
Then the action for the interaction of glueball and heavy-light flavoured
mesons could be derived by combining the gravitational perturbations
in the bulk with the heavy-light meson fields on the flavour branes.
We finally evaluate the decay rates of glueball into two lowest scalar
heavy-light mesons (i.e. the $D^{0}$ meson) by using our effective
action and compare our results with the previous works in \cite{key-25,key-26}.

The organization of this manuscript is as follows. In section 2, we
review the relation of 11d M-theory and type IIA string theory by
their corresponding supergravity. We also give the dynamics of the
glueball from these supergravity systems in holography. In section
3, we briefly review the setup of the flavours in the WSS model and
discuss how to involve the heavy flavour and heavy-light mesons. In
section 4, we detailedly derive the effective action for the interaction
of various glueball and heavy-light meson in holography. In section
5, we identify the lowest heavy-light meson in the model as the lowest
D-meson through some experimental data and we use our effective action
to compute the decay width of the glueball into scalar $D^{0}$ meson.
The final section is our conclusion and summary about this work. At
the end of this manuscript, we collect some essential discussion about
the supergravity, the dynamics of D-brane, the supersymmetry and the
glueball-light meson interaction in Appendix A, B, C respectively
which are also useful to this work.

\section{The dynamics of glueball}

\subsection{The geometry in supergravity}

The WSS model is based on the $\mathrm{AdS_{7}}/\mathrm{CFT_{\mathrm{6}}}$
correspondence obtained by coincident $N_{c}$ M5-branes in 11-dimensional
(11d) M-theory and the dual field theory on the M5-brane is a 6-dimensional
$(0,2)$ superconformal field theory. The effective supergravity description
provides a 11d geometrical near-horizon solution which takes the topology
of $AdS_{7}\times S^{4}$. Let us denote the extended directions of
the M5-branes by $\left\{ X^{\mu},X^{4},X^{11}\right\} $ with indices
$\mu,\nu=0,1,2,3$, so the 11d non-extremal black brane solution is
given as \cite{key-30}\footnote{The equations of motion for the bulk fields can also been found in
the Appendix A.},

\begin{align}
ds_{\left(\mathrm{11d-black}\right)}^{2}= & \frac{r^{2}}{L^{2}}\left[f\left(r\right)\left(dX^{0}\right)^{2}+\delta_{ij}dX^{i}dX^{j}+\left(dX^{4}\right)^{2}+\left(dX^{11}\right)^{2}\right]+\frac{L^{2}}{r^{2}}\frac{dr^{2}}{f\left(r\right)}+\frac{L^{2}}{4}d\Omega_{4}^{2},\nonumber \\
f\left(r\right)= & 1-\frac{r_{KK}^{6}}{r^{6}},\label{eq:1}
\end{align}
where $\Omega_{4},L$ denotes the volume element on $S^{4}$ and the
curvature radius of the $\mathrm{AdS_{7}}$ respectively. The indices
$i,j$ run from 1 to 3 and $r$ represents the radius coordinate of
the transverse directions. Note that this background also contains
a non-vanished Romand-Romand strength $F_{4}$ with $N_{c}$ units
of flux on $S^{4}$. As the 11d M-theory compactified on a cycle is
equivalent to 10d type-IIA string theory, one can obtain non-conformal
D4-branes by the dimensional reduction which preserves circle with

\begin{equation}
X^{11}\sim X^{11}+2\pi R_{11},\ R_{11}=g_{s}l_{s},\ l_{s}^{2}=\alpha^{\prime}.
\end{equation}
Afterwards, the dual field theory becomes a 5d super Yang-Mills theory
on the D4-branes.

In order to obtain a pure Yang-Mills or QCD-like theory, a further
compactification with a double Wick-rotation was proposed by Witten
based on the above holographic duality. Accordingly the metric (\ref{eq:1})
with the double Wick-rotation $X^{0}\rightarrow-iX^{4},X^{4}\rightarrow-iX^{0}$
corresponds to a bubble configuration of D4-brane which is given as
\cite{key-11,key-12},

\begin{align}
ds_{\left(11\mathrm{d}\mathrm{-bubble}\right)}^{2}= & \frac{r^{2}}{L^{2}}\left[\eta_{\mu\nu}dX^{\mu}dX^{\nu}+f\left(r\right)\left(dX^{4}\right)^{2}+\left(dX^{11}\right)^{2}\right]+\frac{L^{2}}{r^{2}}\frac{dr^{2}}{f\left(r\right)}+\frac{L^{2}}{4}d\Omega_{4}^{2},\label{eq:3}
\end{align}
so that the dual field theory exhibits confinements in this geometry.
The supersymmetry of the D4-branes is broken down by compactifying
$X^{4}$ on another circle, hence the fermionic gluinos become massive
at tree level by imposing the antiperiodic boundary conditions and
the adjoint scalars also acquire masses through its loop corrections.
Therefore the gauge bosons are the only degrees of freedom in the
limit of large Kaluza-Klein (KK) mass scale. Employing the relations
of the dimensional reduction, the 10d metric could be written as \cite{key-25,key-26,key-32},

\begin{equation}
ds_{\left(11\mathrm{d}\right)}^{2}=e^{-2\Phi/3}ds_{\left(10\mathrm{d}\right)}^{2}+e^{4\Phi/3}\left(dX^{11}+A_{M}dX^{M}\right)^{2},\label{eq:4}
\end{equation}
where $M=0,1...9$ and $\Phi=\frac{3}{2}\ln\left(\frac{r}{L}\right)$
is the 10d dilaton field. Note that according to the above discussion
we have $A_{m}=0$ and a periodic condition for $X^{4}$,

\begin{equation}
X^{4}\sim X^{4}+2\pi\delta X^{4},\ \delta X^{4}=\frac{1}{M_{KK}}=\frac{L^{2}}{3r_{KK}}.
\end{equation}
The relation of $r_{KK}$ and $M_{KK}$ can be determined by eliminating
the conical singularity at $r=r_{KK}$. Using alternative radial coordinates
$U\in[U_{KK},+\infty)$ and $Z\in[-\infty,+\infty)$ defined as

\begin{equation}
L=2R,\ U=\frac{r^{2}}{2L},\ K\left(Z\right)=1+Z^{2}=\frac{r^{6}}{r_{KK}^{6}}=\frac{U^{3}}{U_{KK^{3}}},\label{eq:6}
\end{equation}
the 10d metric in (\ref{eq:4}) can be explicitly written as,

\begin{align}
ds_{\left(10\mathrm{d}\right)}^{2}= & \left(\frac{U}{R}\right)^{3/2}\left[\eta_{\mu\nu}dX^{\mu}dX^{\nu}+f\left(U\right)\left(dX^{4}\right)^{2}\right]+\left(\frac{R}{U}\right)^{3/2}\left[\frac{dU^{2}}{f\left(U\right)}+U^{2}d\Omega_{4}^{2}\right],\nonumber \\
f\left(U\right)= & 1-\frac{U_{KK}^{3}}{U^{3}},\ e^{\Phi}=\left(\frac{U}{R}\right)^{3/4},\ F_{4}=dC_{3}=\frac{2\pi N_{c}}{V_{4}}\epsilon_{4},\label{eq:7}
\end{align}
where $\epsilon_{4}$ denotes a unit volume element on $S^{4}$ and
$F_{4}$ is the field strength of the Romand-Romand form $C_{3}$.
Note that the metric in (\ref{eq:7}) corresponds to D4 bubble configuration
which is obtained by a double Wick-rotation from the black brane solution.
And the 11th direction $X^{11}$ becomes vanished in the large $N_{c}$
limit. In terms of QCD variables we additionally have the following
relations,

\begin{equation}
\lambda=g_{\mathrm{YM}}^{2}N_{c},\ g_{\mathrm{YM}}^{2}=2\pi g_{s}l_{s}M_{KK},\ R^{3}=\pi g_{s}N_{c}l_{s}^{3},
\end{equation}
where $\lambda,g_{\mathrm{YM}},g_{s},l_{s}$ respectively represents
the 't Hooft coupling constant, Yang-Mills coupling constant, string
coupling constant and the length of string. 

\subsection{Gravitational fluctuations as the glueball field in holography}

Since glueball is the gauge invariant composite state in the Yang-Mills
theory, they holographically corresponds to the gravitational perturbations
of the near-horizon geometry in the supergravity description \cite{key-23,key-24,key-25,key-26,key-27}.
Such metric fluctuations are sourced by the operators in the dual
field theory i.e. the 5d compactified Yang-Mills theory. Hence let
us introduce the gravitational perturbations to the background (\ref{eq:3})
by rewriting the 11d metric as $G_{MN}\rightarrow G_{MN}^{\left(0\right)}+\delta G_{MN}$
where $G_{MN}^{\left(0\right)}$ refers to the metric in (\ref{eq:3}).
And it would be convenient to integrate over the $S^{4}$ in the 11d
supergravity action since $S^{4}$ is not very necessary in the following
discussion. So we can obtain the kinetic action for glueball by imposing
the metric (\ref{eq:3}) into the 11d supergravity action and it leads
to,

\begin{align}
S_{11\mathrm{D}}= & \frac{1}{2\kappa_{11}^{2}}\left(\frac{L}{2}\right)^{4}\Omega_{4}\int d^{7}x\sqrt{-\det G}\left(\mathcal{R}_{11\mathrm{D}}+\frac{30}{L^{2}}\right)\nonumber \\
= & \frac{1}{2\kappa_{11}^{2}}\left(\frac{L}{2}\right)^{4}\Omega_{4}\mathcal{C}_{E,D,T}S_{G},\label{eq:9}
\end{align}
where $S_{G}$ denotes the kinetic term of the glueball and $\mathcal{C}_{E,D,T}$
is a constant determined by the various gravitational fluctuations.
Note that the pre-factor in the above action has to be normalized
as $\frac{1}{2\kappa_{11}^{2}}\left(\frac{L}{2}\right)^{4}\Omega_{4}\mathcal{C}_{E,D,T}=1$
in which $\mathcal{C}_{E,D,T}$ is completely fixed. We will use this
action to discuss the dynamics of the glueball with various gravitational
perturbations in the the rest of the manuscript.

\subsubsection*{A. The exotic scalar glueball}

The lowest exotic scalar glueball has quantum number $J^{CP}=0^{++}$
which corresponds to the exotic polarizations of the bulk graviton.
The 11d components of $\delta G_{MN}$ are given as \cite{key-26},

\begin{eqnarray}
\delta G_{44} & = & -\frac{r^{2}}{L^{2}}f\left(r\right)H_{E}\left(r\right)G_{E}\left(x\right),\nonumber \\
\delta G_{\mu\nu} & = & \frac{r^{2}}{L^{2}}H_{E}\left(r\right)\left[\frac{1}{4}\eta_{\mu\nu}-\left(\frac{1}{4}+\frac{3r_{KK}^{6}}{5r^{6}-2r_{KK}^{6}}\right)\frac{\partial_{\mu}\partial_{\nu}}{M_{E}^{2}}\right]G_{E}\left(x\right),\nonumber \\
\delta G_{11,11} & = & \frac{r^{2}}{4L^{2}}H_{E}\left(r\right)G_{E}\left(x\right),\nonumber \\
\delta G_{rr} & = & -\frac{L^{2}}{r^{2}}\frac{1}{f\left(r\right)}\frac{3r_{KK}^{6}}{5r^{6}-2r_{KK}^{6}}H_{E}\left(r\right)G_{E}\left(x\right),\nonumber \\
\delta G_{r\mu} & = & \frac{90r^{7}r_{KK}^{6}}{M_{E}^{2}L^{2}\left(5r^{6}-2r_{KK}^{6}\right)^{2}}H_{E}\left(r\right)\partial_{\mu}G_{E}\left(x\right),\label{eq:10}
\end{eqnarray}
where the eigenvalue equation for function $H_{E}\left(r\right)$
is,

\begin{equation}
\frac{1}{r^{3}}\frac{d}{dr}\left[r\left(r^{6}-r_{KK}^{6}\right)\frac{d}{dr}H_{E}\left(r\right)\right]+\left[\frac{432r^{2}r_{KK}^{12}}{\left(5r^{6}-2r_{KK}^{6}\right)^{2}}+L^{4}M_{E}^{2}\right]H_{E}\left(r\right)=0.\label{eq:11}
\end{equation}
Note that the above components in (\ref{eq:10}) involve asymptotics
in the bulk as $\delta G_{44}=-4\delta G_{11}=-4\delta G_{22}=-4\delta G_{33}=-4\delta G_{11,11}$
for $r\rightarrow\infty$. Imposing the metric (\ref{eq:3}) with
fluctuations (\ref{eq:10}) and the eigenvalue equation for function
$H_{E}\left(r\right)$ into action (\ref{eq:9}), we obtain the kinetic
term of the exotic scalar glueball which is,

\begin{equation}
S_{G_{E}\left(x\right)}=-\frac{1}{2}\int d^{4}x\left[\left(\partial_{\mu}G_{E}\right)^{2}+M_{E}^{2}G_{E}^{2}\right],\label{eq:12}
\end{equation}
where

\begin{equation}
\mathcal{C}_{E}=\int_{r_{KK}}^{\infty}dr\frac{r^{3}}{L^{3}}\frac{5}{8}H_{E}^{2}\left(r\right).
\end{equation}

\subsubsection*{B. The dilatonic and tensor glueball}

The scalar glueball of $0^{++}$ mode corresponds to the fluctuations
of the metric as \cite{key-26},

\begin{align}
\delta G_{11,11} & =-3\frac{r^{2}}{L^{2}}H_{D}\left(r\right)G_{D}\left(x\right),\nonumber \\
\delta G_{\mu\nu} & =\frac{r^{2}}{L^{2}}H_{D}\left(r\right)\left[\eta^{\mu\nu}-\frac{\partial^{\mu}\partial^{\nu}}{M_{D}^{2}}\right]G_{D}\left(x\right).\label{eq:14}
\end{align}
We refer to the upon mode as ``dilatonic'' since $\delta G_{11,11}$
reduces to the 10d dilaton.

The tensor glueball with $J^{CP}=2^{++}$ corresponds to the metric
fluctuations which includes a transverse traceless polarization. A
choice of the non-vanishing components of the graviton polarizations
could be,

\begin{equation}
\delta G_{\mu\nu}=-\frac{r^{2}}{L^{2}}H_{T}\left(r\right)T_{\mu\nu}\left(x\right),\label{eq:15}
\end{equation}
where $T_{\mu\nu}\equiv\mathcal{T}_{\mu\nu}G_{T}\left(x\right)$.
$\mathcal{T}_{\mu\nu}$ is a constant symmetric tensor normalized
as $\mathcal{T}_{\mu\nu}\mathcal{T}^{\mu\nu}=1$ and satisfies the
traceless condition $\eta^{\mu\nu}\mathcal{T}_{\mu\nu}=0$. The eigenvalue
equation for functions $H_{D,T}\left(r\right)$ are given as,

\begin{equation}
\frac{1}{r^{3}}\frac{d}{dr}\left[r\left(r^{6}-r_{KK}^{6}\right)\frac{d}{dr}H_{D,T}\left(r\right)\right]+L^{4}M_{D,T}^{2}H_{D,T}\left(r\right)=0.\label{eq:16}
\end{equation}
Plugging metric (\ref{eq:3}) with fluctuations (\ref{eq:14}) and
(\ref{eq:15}) into action (\ref{eq:9}) respectively leads to the
dynamics of the dilatonic scalar and tensor glueball,

\begin{align}
S_{G_{D}\left(x\right)} & =-\frac{1}{2}\int d^{4}x\left[\left(\partial_{\mu}G_{D}\right)^{2}+M_{D}^{2}G_{D}^{2}\right],\nonumber \\
S_{T\left(x\right)} & =-\frac{1}{4}\int d^{4}x\left[T_{\mu\nu}\left(\partial^{2}-M_{T}^{2}\right)T^{\mu\nu}\right],
\end{align}
where the corresponding normalization constant is,

\begin{align}
\mathcal{C}_{D} & =6\int_{r_{KK}}^{\infty}dr\frac{r^{3}}{L^{3}}H_{D}^{2}\left(r\right),\nonumber \\
\mathcal{C}_{T} & =2\int_{r_{KK}}^{\infty}dr\frac{r^{3}}{L^{3}}H_{T}^{2}\left(r\right).
\end{align}

\section{The flavours and mesons}

\subsection{The light flavour and meson}

There are also flavours in this model which are introduced by embedding
a stack of $N_{f}$ probe D8- and anti-D8-branes ($\mathrm{D}8/\overline{\mathrm{D}8}$-brane)
at the antipodal position of the background geometry. The configurations
of the various branes are illustrated in Table \ref{tab:1}. 

\begin{table}[h]
\begin{centering}
\begin{tabular}{|c|c|c|c|c|c|c|c|c|c|c|}
\hline 
 & 0 & 1 & 2 & 3 & $\left(4\right)$ & 5$\left(U\right)$ & 6 & 7 & 8 & 9\tabularnewline
\hline 
\hline 
$N_{c}\ \mathrm{D4}$  & - & - & - & - & - &  &  &  &  & \tabularnewline
\hline 
$N_{f}\ \mathrm{D}8/\overline{\mathrm{D}8}$ & - & - & - & - &  & - & - & - & - & -\tabularnewline
\hline 
\end{tabular}
\par\end{centering}
\caption{\label{tab:1}The D-brane configurations: \textquotedblleft -\textquotedblright{}
denotes the world volume directions of the D-branes.}
\end{table}

\begin{figure}
\begin{centering}
\includegraphics[scale=0.5]{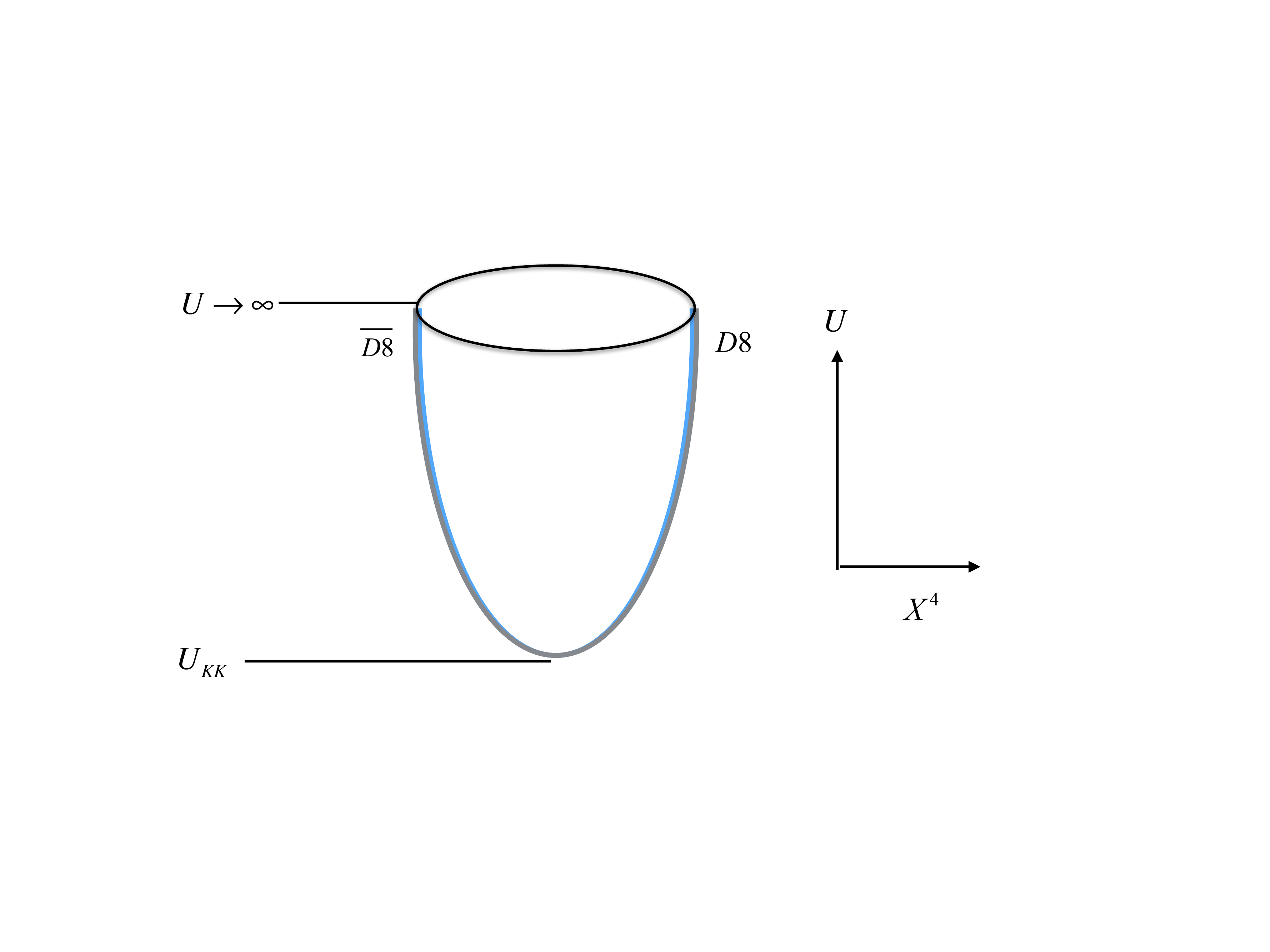}
\par\end{centering}
\caption{\label{fig:1}The D-brane configuration in the $X^{4}-U$ plane where
$X^{4}$ is compactified on $S^{1}$. The bubble background (cigar)
is produced by $N_{c}$ D4-branes . The $N_{f}$ light flavour $\mathrm{D}8/\overline{\mathrm{D}8}$-branes
(L) living at antipodal position of the cigar are represented by the
blue line. The $\mathrm{D}8/\overline{\mathrm{D}8}$-branes are connected
to each other at the bottom of the bulk ($U=U_{KK}$).}

\end{figure}

The $N_{f}$ flavoured fermions arise from the lowest modes of the
$4-8$ strings\footnote{We use ``$p-q$ string'' and ``$p-\bar{q}$ string'' to denote
the open string with one end attached to the $\mathrm{D}p$-brane
and the other end to the $\mathrm{D}q$-brane and $\overline{\mathrm{D}q}$-brane
respectively. For example, ``$4-\bar{8}$ string'' refers to the
open string with one end attached to the $\mathrm{D}4$-brane and
the other end to the $\overline{\mathrm{D}8}$-brane.} and $4-\bar{8}$ strings which belong to the fundamental representation
of the colour group $U\left(N_{c}\right)$ and flavour group $U\left(N_{f}\right)$.
In order to define the chirality of these fundamental fermions, a
natural choice is to perform the GSO (Gliozzi-Scherk-Olive) projection
\cite{key-12,key-33}. Since the GSO projection for $4-8$ strings
is opposite to the projection chosen for the $4-\bar{8}$ strings,
these fundamental fermions could be interpreted as quarks with opposite
chiralities in QCD. And the gauge symmetry introduced by $\mathrm{D}8/\overline{\mathrm{D}8}$-branes
can be therefore denoted as $U\left(N_{f}\right)_{L}\times U\left(N_{f}\right)_{R}$,
namely the chiral symmetry. Note that the $4-8/\bar{8}$ strings has
zero length in this D4/D8 configuration, the fundamental fermions
are therefore massless which correspond accordingly to the light flavoured
quarks in QCD. In this sense it motivates us to name these flavour
branes as ``light flavour branes''. As shown in Table \ref{tab:1},
the quarks and anti-quarks live on the separate position of $X^{4}$.
However the global chiral symmetry $U\left(N_{f}\right)_{L}\times U\left(N_{f}\right)_{R}$
is broken spontaneously since the geometry of the D4 bubble configuration
forces the $\mathrm{D}8/\overline{\mathrm{D}8}$-brane to be connected
at the bottom of the bulk as illustrated in Figure\ref{fig:1}.

Thus the induced metric on the flavour branes could be straightforward
obtained by putting $\mathrm{D}8/\overline{\mathrm{D}8}$-branes onshell
i.e. imposing the condition $\frac{dx^{4}}{dU}=0$, then we have,

\begin{equation}
ds_{\mathrm{D}8}^{2}=g_{MN}dX^{M}dX^{N}=\left(\frac{U}{R}\right)^{3/2}\eta_{\mu\nu}dx^{\mu}dx^{\nu}+\left(\frac{R}{U}\right)^{3/2}\left[\frac{dU^{2}}{f\left(U\right)}+U^{2}d\Omega_{4}^{2}\right],\label{eq:19}
\end{equation}
where the indices $M,N$ run over the flavour brane. So the action
of the joined light flavour branes describes the dynamics of $q\bar{q}$
mesons, which contains the light quarks only, through the flavoured
gauge fields on the worldvolume of the $\mathrm{D}8/\overline{\mathrm{D}8}$-branes.
The explicit form of the action is,

\begin{equation}
S_{\mathrm{D8}}=S_{\mathrm{DBI}}+S_{\mathrm{CS}},
\end{equation}
where\footnote{For $N_{f}$ D8-branes, all the actions in (\ref{eq:21}) should include
a pre-factor $N_{f}$. Nonetheless we do not pick it up because a
pre-factor does not affect any results.}

\begin{align}
S_{\mathrm{DBI}}= & -T_{8}\mathrm{Tr}\int_{\mathrm{D}8/\overline{\mathrm{D}8}}d^{9}xe^{-\Phi}\sqrt{-\det\left(g+2\pi\alpha^{\prime}F\right)}\nonumber \\
= & -T_{8}\mathrm{Tr}\int_{\mathrm{D}8/\overline{\mathrm{D}8}}d^{9}xe^{-\Phi}\sqrt{-g}\left[1+\frac{1}{4}\left(2\pi\alpha^{\prime}\right)^{2}g^{MN}g^{KL}F_{MK}F_{NL}+\mathcal{O}\left(F^{4}\right)\right],\nonumber \\
S_{\mathrm{CS}}= & \mu_{8}\int_{\mathrm{D}8/\overline{\mathrm{D}8}}C_{3}\wedge\left(2\pi\alpha^{\prime}\right)^{3}\mathrm{Tr}\left(F\wedge F\wedge F\right).\label{eq:21}
\end{align}
Here $g$ and $\Phi$ is the induced metric and dilaton given in (\ref{eq:19})
(\ref{eq:7}) respectively. $F$ refers to the field strength of the
flavoured gauge field $A_{M}$ on the D8-branes which is defined as
$F_{MN}=\partial_{M}A_{N}-\partial_{N}A_{M}+\left[A_{M},A_{N}\right]$\footnote{In this notation, the gauge field $A_{M}$ is anti-Hermitian which
means $A_{M}^{\dagger}=-A_{M}$ and $F_{MN}^{\dagger}=-F_{MN}$.}. Note that in this model the non-zero components of the gauge field
$A_{M}$ are only $\left\{ A_{\mu},A_{Z}\right\} $ which are functions
dependent on the coordinates $\left\{ x^{\mu},Z\right\} $. Since
the flavour branes are probes, we could obtain the following 5d quadratic
action after imposing the solution (\ref{eq:19}) (\ref{eq:7}) to
(\ref{eq:21}) and integrating over $S^{4}$ which is,

\begin{equation}
S_{\mathrm{DBI}}^{\mathrm{YM}}=-\kappa\mathrm{Tr}\int d^{4}x\int_{-\infty}^{+\infty}dZ\left[\frac{1}{2}K\left(Z\right)^{-1/3}\eta^{\mu\rho}\eta^{\nu\sigma}F_{\mu\nu}F_{\rho\sigma}+K\left(Z\right)M_{KK}^{2}\eta^{\mu\nu}F_{\mu Z}F_{\nu Z}\right],\label{eq:22}
\end{equation}
where

\begin{equation}
\kappa=\frac{1}{3}\left(2\pi\alpha^{\prime}\right)^{2}T_{8}g_{s}^{-1}\omega_{4}R^{9/2}U_{KK}^{1/2}=\frac{\lambda N_{c}}{216\pi^{3}},\ \omega_{4}=\frac{8\pi^{2}}{3},\ M_{KK}^{2}=\frac{9U_{KK}}{4R^{3}}.
\end{equation}
The light flavour meson fields are identified as the expansion modes
of $A_{M}$ by a complete sets $\left\{ \psi_{n}\right\} $ in the
model. Explicitly $A_{M}$ is assumed to be expanded as \footnote{In (\ref{eq:22}) there is a residual gauge symmetry in the action.
So it would be convenient to choose a gauge which keeps $A_{Z}$ nonzero
for our purposes. The frequently chosen gauge which keeps $A_{Z}=0$
leads to a different parametrization of the Goldstone bosons in this
model but physically equivalent situation.},

\begin{equation}
A_{\mu}=\sum_{n=1}B_{\mu}^{\left(n\right)}\left(x\right)\psi_{n}\left(Z\right),\ A_{Z}=\varphi^{\left(0\right)}\left(x\right)\psi_{0}^{\prime}\left(Z\right)+\sum_{n=1}m_{n}^{-1}\varphi^{\left(n\right)}\left(x\right)\psi_{n}^{\prime}\left(Z\right).\label{eq:24}
\end{equation}
so the lowest mode of $A_{\mu}$ and $A_{Z}$ could be interpreted
as the lightest vector $\rho$ meson and scalar $\pi$ meson. Therefore
if we identify $B_{\mu}^{\left(1\right)}\left(x\right)=\rho_{\mu}\left(x\right),\varphi^{\left(0\right)}\left(x\right)=U_{KK}\pi\left(x\right)$
with the normalization, 

\begin{equation}
2\kappa\int_{-\infty}^{+\infty}dZK\left(Z\right)^{-1/3}\left[\psi_{1}\left(Z\right)\right]^{2}=1,\ \ 2\kappa\left(U_{KK}M_{KK}\right)^{2}\int_{-\infty}^{+\infty}dZK\left(Z\right)\left[\psi_{0}^{\prime}\left(Z\right)\right]^{2}=1,
\end{equation}
the effective 4d action of $\pi$ and $\rho$ meson denoted by $S_{\pi-\rho}^{\mathrm{eff}}$
can be obtained from (\ref{eq:22}) whose form is\footnote{The dynamics of $\varphi^{\left(1\right)}$ is absorbed by the $\rho$
meson through a gauge transformation $B_{\mu}^{\left(1\right)}\rightarrow B_{\mu}^{\left(1\right)}+m_{n}^{-1}\partial_{\mu}\varphi^{\left(1\right)}$. },

\begin{equation}
S_{\pi-\rho}^{\mathrm{eff}}=-\mathrm{Tr}\int d^{4}x\left[\frac{1}{2}\left(\partial_{\mu}\pi\right)^{2}+\frac{1}{4}\partial_{[\mu}\rho_{\nu]}\partial^{[\mu}\rho^{\nu]}+\frac{1}{2}\lambda_{1}M_{KK}^{2}\rho_{\mu}^{2}+...\left(\mathrm{interactions}\right)\right].\label{eq:26}
\end{equation}
The Chern-Simons (CS) term in (\ref{eq:21}) corresponds to the Wess-Zumino-Witten
(WZW) term in chiral QCD. However we will not discuss its explicit
form in this manuscript because it is independent on the metric thus
does not include the interaction of meson and glueball.

\subsection{The heavy flavour and heavy-light flavoured meson}

\begin{figure}
\begin{centering}
\includegraphics[scale=0.5]{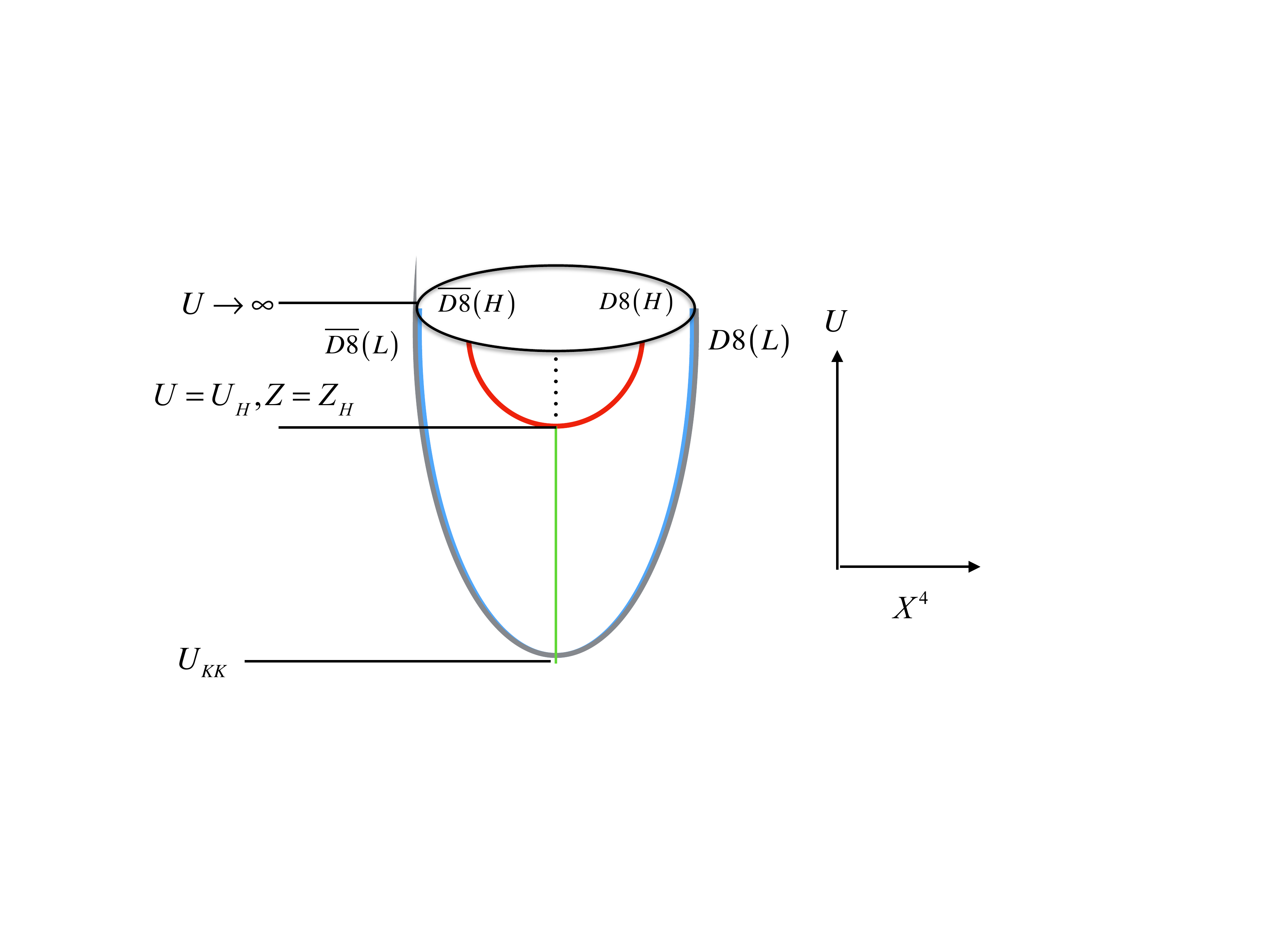}
\par\end{centering}
\caption{\label{fig:2}The D-brane configuration involving the heavy flavour
in the $X^{4}-U$ plane. Additional to the light flavour branes,  a
pair of heavy flavour $\mathrm{D}8/\overline{\mathrm{D}8}$-brane
(H) (denoted by the red line) is introduced in to the D4/D8 configuration
which is separated from the light flavour branes. The heavy flavour
$\mathrm{D}8/\overline{\mathrm{D}8}$-branes are connected to each
other at $U=U_{H}>U_{KK}$ and the massive multiplets are produced
by the heavy-light string (HL-string denoted by the green line) stretched
between the light and heavy flavour branes which is denoted by the
green line in this figure.}
\end{figure}

The heavy flavour could be introduced into this model by considering
an extra pair of flavoured $\mathrm{D}8/\overline{\mathrm{D}8}$-brane
which is separated from the other $N_{f}$ (light-flavoured) $\mathrm{D}8/\overline{\mathrm{D}8}$-branes
with an open string (the heavy-light string) stretched between them
\cite{key-28,key-29,key-34,key-35} as illustrated in Figure \ref{fig:2}.
According to string theory, the heavy-light (HL) string stretched
between the separated branes creates additional multiplets which could
be approximated near the worldvolume of the light flavour branes by
local vector fields. Due to the finite separation of the heavy and
light flavour branes, the HL-string has a non-zero length so that
those multiplets acquire mass through the non-zero vacuum expectation
value (VEV) of the HL-string. Hence the multiplets could be interpreted
as the heavy-flavoured mesons with massive quarks. We note that this
mechanism to acquire the massive fields is exactly the ``Higgs mechanism''
in string theory \cite{key-30}.

In order to include HL-multiple states in the actions (\ref{eq:21}),
we need to replace the gauge fields on the light flavour branes by
its matrix-valued formula according to string theory,

\begin{equation}
A_{M}\rightarrow\boldsymbol{\mathrm{A}}_{M}=\left(\begin{array}{cc}
A_{M} & \Phi_{M}\\
-\Phi_{M}^{\dagger} & 0
\end{array}\right).\label{eq:27}
\end{equation}
In our setup $\boldsymbol{\mathrm{A}}_{M}$ is $\left(N_{f}+1\right)\times\left(N_{f}+1\right)$
matrix-valued 1-form while $A_{M}$ is $N_{f}\times N_{f}$ valued
1-form. And $\Phi_{M}$ is $N_{f}\times1$ valued vector. Thus the
field strength of $\boldsymbol{\mathrm{A}}_{M}$ becomes a matrix-valued
2-form as,

\begin{equation}
F_{MN}\rightarrow\boldsymbol{\mathrm{F}}_{MN}=\left(\begin{array}{cc}
F_{MN}-\Phi_{[M}\Phi_{N]}^{\dagger} & \partial_{[M}\Phi_{N]}+A_{[M}\Phi_{N]}\\
-\partial_{[M}\Phi_{N]}^{\dagger}-\Phi_{[M}^{\dagger}A_{N]} & -\Phi_{[M}^{\dagger}\Phi_{N]}
\end{array}\right).\label{eq:28}
\end{equation}
And we have assumed that the non-zero components of $\Phi_{M}$ are
$\left\{ \Phi_{\mu},\Phi_{Z}\right\} $ which depends only on $\left\{ x^{\mu},Z\right\} $
in order to be compatible with $A_{M}$. Plugging (\ref{eq:27}) (\ref{eq:28})
into the quadratic action (\ref{eq:22}), we obtain the action as,

\begin{align}
S_{\mathrm{DBI}}^{\mathrm{YM}}= & -\frac{1}{4}\left(2\pi\alpha^{\prime}\right)^{2}T_{8}\mathrm{Tr}\int_{\mathrm{D}8/\overline{\mathrm{D}8}}d^{9}xe^{-\Phi}\sqrt{-g}g^{MN}g^{KL}\boldsymbol{\mathrm{F}}_{MK}\boldsymbol{\mathrm{F}}_{NL}\label{eq:29}\\
= & -\kappa\mathrm{Tr}\int d^{4}x\int_{-\infty}^{+\infty}dZ\left[\frac{1}{2}K\left(Z\right)^{-1/3}\eta^{\mu\rho}\eta^{\nu\sigma}\boldsymbol{\mathrm{F}}_{\mu\nu}\boldsymbol{\mathrm{F}}_{\rho\sigma}+K\left(Z\right)M_{KK}^{2}\eta^{\mu\nu}\boldsymbol{\mathrm{F}}_{\mu Z}\boldsymbol{\mathrm{F}}_{\nu Z}\right].\nonumber \\
= & -\kappa\mathrm{Tr}\int d^{4}x\int_{-\infty}^{+\infty}dZ\bigg\{\frac{1}{2}K\left(Z\right)^{-1/3}\left[\left(F_{\mu\nu}-a_{\mu\nu}\right)\left(F^{\mu\nu}-a^{\mu\nu}\right)-2f_{\mu\nu}^{\dagger}f^{\mu\nu}+b_{\mu\nu}b^{\mu\nu}\right]\nonumber \\
 & +K\left(Z\right)M_{KK}^{2}\eta^{\mu\nu}\left[\left(F_{\mu Z}-a_{\mu Z}\right)\left(F_{\nu Z}-a_{\nu Z}\right)-2f_{\mu Z}^{\dagger}f_{\nu Z}+b_{\mu Z}b_{\nu Z}\right]\bigg\},\label{eq:30}
\end{align}
where ``$\mathrm{Tr}$'' refers to the ``flavour trace'' and

\begin{align}
f_{MN}^{\dagger}=-\partial_{[M}\Phi_{N]}^{\dagger}-\Phi_{[M}^{\dagger}A_{N]}, & \ \ \ \ b_{MN}=\Phi_{[M}^{\dagger}\Phi_{N]},\nonumber \\
f_{MN}=\partial_{[M}\Phi_{N]}+A_{[M}\Phi_{N]}, & \ \ \ \ a_{MN}=\Phi_{[M}\Phi_{N]}^{\dagger}.
\end{align}
Note that the DBI action has to involve the transverse modes of the
brane since the HL-string with finite separation provides a non-zero
VEV in our setup. For the reader\textquoteright s convenience, we
have summarized the essential derivations to obtain the quadratic
action in (\ref{eq:21}) from the complete DBI action with non-Abelian
excitation in the Appendix B. Then let us denote the only one transverse
``coordinate'' as $\Psi$ for D8-branes and its T-dualitized action
is given as,

\begin{equation}
S_{\Psi}=-\tilde{T}_{8}\mathrm{Tr}\int_{\mathrm{D}8/\overline{\mathrm{D}8}}d^{9}xe^{-\Phi}\sqrt{-\det g}\left\{ \frac{1}{2}D_{M}\Psi D^{M}\Psi+\frac{1}{4}\left[\Psi,\Psi\right]^{2}\right\} ,\label{eq:32}
\end{equation}
with $D_{M}\Psi=\partial_{M}\Psi+\left[\boldsymbol{\mathrm{A}}_{M},\Psi\right]$
and $\tilde{T}_{8}=\left(2\pi\alpha^{\prime}\right)^{2}T_{8}$.  By
the extrema of the potential contribution or $\left[\Psi,\left[\Psi,\Psi\right]\right]=0$,
we can define the moduli solution of $\Psi$ for $N_{f}$ light branes
separated from one heavy brane with a finite VEV $v$ as \cite{key-30,key-31},

\begin{equation}
\Psi=\left(\begin{array}{cc}
-\frac{v}{N_{f}}\boldsymbol{1}_{N_{f}} & 0\\
0 & v
\end{array}\right).\label{eq:33}
\end{equation}
By using the solution (\ref{eq:33}), the action (\ref{eq:32}) could
be rewritten as,

\begin{equation}
S_{\Psi}=-\tilde{T}_{8}v^{2}\frac{\left(N_{f}+1\right)^{2}}{N_{f}^{2}}\mathrm{Tr}\int d^{4}x\int_{-\infty}^{+\infty}dZe^{-\Phi}\sqrt{-\det g}g^{MN}\Phi_{M}^{\dagger}\Phi_{N},\label{eq:34}
\end{equation}
We note here the total action of the light flavour branes involving
the HL fields is collected as,

\begin{equation}
S_{\mathrm{D8}}=S_{\mathrm{DBI}}^{\mathrm{YM}}+S_{\Psi}+S_{\mathrm{CS}},
\end{equation}
where $S_{\mathrm{CS}}$ takes the same formula as (\ref{eq:21})
but replacing $F\rightarrow\boldsymbol{\mathrm{F}}$. 

Along the discussion in the sector of light flavour , we further assume
that the HL-multiplet $\Phi_{M}$ could be expanded as\footnote{We assume that $\Phi_{Z}$ is expanded with a pre-factor $U_{KK}$
so that the base functions $\tilde{\phi}_{m}\left(Z\right)$'s are
all dimensionless. },

\begin{align}
\Phi_{\mu}=\sum_{m}Q_{\mu,m}\left(x\right)\phi_{m}\left(Z\right), & \ \ \ \ \Phi_{Z}=U_{KK}\sum_{m}Q_{m}\left(x\right)\tilde{\phi}_{m}\left(Z\right),\nonumber \\
\Phi_{\mu}^{\dagger}=\sum_{m}Q_{\mu,m}^{\dagger}\left(x\right)\phi_{m}\left(Z\right), & \ \ \ \ \Phi_{Z}^{\dagger}=U_{KK}\sum_{m}Q_{m}^{\dagger}\left(x\right)\tilde{\phi}_{m}\left(Z\right),\label{eq:36}
\end{align}
with the normalization conditions

\begin{align}
4\kappa\int_{-\infty}^{+\infty}dZK\left(Z\right)^{-1/3}\phi_{m}\left(Z\right)\phi_{n}\left(Z\right) & =\delta_{mn},\nonumber \\
4\kappa M_{KK}^{2}U_{KK}^{2}\int_{-\infty}^{+\infty}dZK\left(Z\right)\tilde{\phi}_{m}\left(Z\right)\tilde{\phi}_{n}\left(Z\right) & =\delta_{mn},\nonumber \\
4\kappa\int_{-\infty}^{+\infty}dZK\left(Z\right)\phi_{m}^{\prime}\left(Z\right)\phi_{n}^{\prime}\left(Z\right) & =\Lambda_{n}\delta_{mn},\label{eq:37}
\end{align}
then we can obtain the kinetic term of the massive HL-fields from
(\ref{eq:30}) (\ref{eq:34}) which is,

\begin{align}
S_{\mathrm{HL}}= & -\mathrm{Tr}\int d^{4}x\sum_{n}\bigg[\frac{1}{2}\partial_{\mu}Q_{n}^{\dagger}\partial^{\mu}Q_{n}+\frac{1}{2}M_{S}^{2}Q_{n}^{\dagger}Q_{n}\nonumber \\
 & +\frac{1}{4}\partial_{[\mu}Q_{\nu],n}^{\dagger}\partial^{[\mu}Q_{\ ,n}^{\nu]}+\frac{1}{2}\left(\Lambda_{n}M_{KK}^{2}+M_{V}^{2}\right)Q_{\mu,n}^{\dagger}Q_{\ ,n}^{\mu}\bigg],\label{eq:38}
\end{align}
where
\begin{align}
M_{V}^{2}=v_{V}^{2}\kappa\int_{-\infty}^{+\infty}dZK\left(Z\right)^{1/6}\phi_{m}\left(Z\right)\phi_{n}\left(Z\right), & \ v_{V}^{2}=\frac{4}{27}\frac{\left(1+N_{f}\right)^{2}}{N_{f}^{2}}v^{2}M_{KK}^{3}R^{3}\nonumber \\
M_{S}^{2}=v_{S}^{2}\kappa\int_{-\infty}^{+\infty}dZK\left(Z\right)^{3/2}\tilde{\phi}_{m}\left(Z\right)\tilde{\phi}_{n}\left(Z\right), & \ v_{S}^{2}=\frac{64}{2187}\frac{\left(1+N_{f}\right)^{2}}{N_{f}^{2}}v^{2}M_{KK}^{9}R^{9}.\label{eq:39}
\end{align}
Note that $M_{V},M_{S}$ relate to the mass term of the heavy-light
meson fields while $M_{V},M_{S}$ might be theoretically divergent
since they are defined by an integral (\ref{eq:39}). To figure out
this issue, we could choose the physical mass as the value of $M_{V},M_{S}$
which implies that $M_{V},M_{S}$ have to be renormalized by the holographic
counterterms on the D-brane. The holographic counterterms in this
model have been discussed in, such as \cite{key-17,key-21}\footnote{A more general discussion about the holographic renormalization on
probe D-brane can be reviewed in \cite{key-36,key-37}. }, so briefly speaking, as an effective theory of meson, we may keep
the finite part in the integral (\ref{eq:39}) as the physical mass
if $M_{V},M_{S}$ are divergent.

\section{The interactions of glueball and heavy-light flavoured mesons}

Since the gravitational perturbation signals the glueball states in
holography, let us include the gravitational fluctuations in this
section in order to obtain the effective Lagrangian of glueball and
mesons. The glueball field could be involved by replacing the metric
by $G\rightarrow G+\delta G$ as we discussed in Section 1, the interaction
of glueball and various mesons will therefore arise once we add the
gravitational fluctuations to the terms depended on the metric in
the action. Accordingly it only concerns the Yang-Mills action (\ref{eq:28})
and the mass term (\ref{eq:34}) since the Chern-Simons action in
(\ref{eq:21}) is independent on the metric. Thus let us first rewrite
the Yang-Mills action (\ref{eq:28}) as,

\begin{equation}
S_{\mathrm{DBI}}^{\mathrm{YM}}\equiv\sum_{i=1}^{3}S_{i}\equiv S_{\mathrm{L}}+S_{\mathrm{HL}}+\sum_{i=1}^{3}\left(S_{i}^{HL-L}+S_{i}^{L-G}+S_{i}^{HL-G}+S_{i}^{HL-L-G}\right),\label{eq:40}
\end{equation}
where $S_{\mathrm{L}},S_{\mathrm{HL}}$ refers to the kinetic terms
of light and HL-mesons given respectively in (\ref{eq:26}) (\ref{eq:38})
and

\begin{align}
S_{1} & =-\frac{1}{4}\left(2\pi\alpha^{\prime}\right)^{2}T_{8}\omega_{4}\mathrm{Tr}\int d^{4}xdZe^{-\Phi}\sqrt{-g_{(5)}}g^{\mu\rho}g^{\nu\sigma}\bigg[\left(F_{\mu\nu}-a_{\mu\nu}\right)\left(F_{\rho\sigma}-a_{\rho\sigma}\right)-2f_{\mu\nu}^{\dagger}f_{\rho\sigma}\nonumber \\
 & \ \ \ +b_{\mu\nu}b_{\rho\sigma}\bigg],\nonumber \\
S_{2} & =-\frac{1}{2}\left(2\pi\alpha^{\prime}\right)^{2}T_{8}\omega_{4}\mathrm{Tr}\int d^{4}xdZe^{-\Phi}\sqrt{-g_{(5)}}\left(g^{\mu\nu}g^{ZZ}-g^{\mu Z}g^{\nu Z}\right)\bigg[\left(F_{\mu Z}-a_{\mu Z}\right)\left(F_{\nu Z}-a_{\nu Z}\right)\nonumber \\
 & \ \ \ -2f_{\mu Z}^{\dagger}f_{\nu Z}+b_{\mu Z}b_{\nu Z}\bigg],\nonumber \\
S_{3} & =-\left(2\pi\alpha^{\prime}\right)^{2}T_{8}\omega_{4}\mathrm{Tr}\int d^{4}xdZe^{-\Phi}\sqrt{-g_{(5)}}g^{\mu\rho}g^{\nu Z}\bigg[\left(F_{\mu\nu}-a_{\mu\nu}\right)\left(F_{\rho Z}-a_{\rho Z}\right)-2f_{\mu\nu}^{\dagger}f_{\rho Z}\nonumber \\
 & \ \ \ +b_{\mu\nu}b_{\rho Z}\bigg].\label{eq:41}
\end{align}
The coupling terms of light mesons and glueball, HL-mesons and glueball,
HL- and light meson, light and HL-meson with glueball have been respectively
denoted by using the index ``L-G'' ``HL-G'' ``HL-L'' ``HL-L-G''
in the action (\ref{eq:40}). Note that only the linear terms of the
glueball field are valid since the gravitational fluctuations are
perturbations in the supergravity description. Then the 10d metric
involving the 11d fluctuations is given as,

\begin{align}
g_{\mu\nu} & =\frac{r^{3}}{L^{3}}\left[\left(1+\frac{L^{2}}{2r^{2}}\delta G_{11,11}\right)\eta_{\mu\nu}+\frac{L^{2}}{r^{2}}\delta G_{\mu\nu}\right],\nonumber \\
g_{44} & =\frac{r^{3}f}{L^{3}}\left[1+\frac{L^{2}}{2r^{2}}\delta G_{11,11}+\frac{L^{2}}{r^{2}f}\delta G_{44}\right],\nonumber \\
g_{rr} & =\frac{L}{rf}\left(1+\frac{L^{2}}{2r^{2}}\delta G_{11,11}+\frac{r^{2}f}{L^{2}}\delta G_{rr}\right),\nonumber \\
g_{r\mu} & =\frac{r}{L}\delta G_{r\mu},\ \ g_{\Omega\Omega}=\frac{r}{L}\left(\frac{L}{2}\right)^{2}\left(1+\frac{L^{2}}{2r^{2}}\delta G_{11,11}\right),\label{eq:42}
\end{align}
with the dilaton,

\begin{equation}
e^{4\Phi/3}=\frac{r^{2}}{L^{2}}\left(1+\frac{L^{2}}{r^{2}}\delta G_{11,11}\right).\label{eq:43}
\end{equation}
 We would explicitly derive the effective action for glueball and
mesons by considering the various gravitational fluctuations $\delta G$.

\subsection{Involving the exotic scalar glueball }

Let us start with the 10d metric (\ref{eq:7}) involving the exotic
polarizations of the bulk graviton given in (\ref{eq:10}). Substituting
(\ref{eq:42}) (\ref{eq:43}) with (\ref{eq:10}) (\ref{eq:36}) into
action (\ref{eq:41}), we can obtain the coupling terms of the glueball
and mesons. Although the calculation is very straightforward, the
result might be a little messy. Let us keep the quadratic terms of
$\Phi_{M}$ as the effective action (\ref{eq:41}), so the interaction
terms of glueball and meson are collected as follows\footnote{We will use $G_{E,D},T$ to denote the interaction involving the exotic,
dilatonic scalar glueball and tensor glueball respectively. The action
for the interaction of glueball and $\pi,\rho$ meson in this model
is collected in Appendix C.},

\begin{align}
S_{1}^{HL-L}= & \mathrm{Tr}\sum_{m,n}\int d^{4}xdZ\mathcal{A}_{1}\eta^{\mu\rho}\eta^{\nu\sigma}\mathcal{I}_{\mu\nu\rho\sigma},\label{eq:44}\\
S_{1}^{L-G_{E}}= & \mathrm{Tr}\int d^{4}xdZ\left(\frac{\mathcal{B}_{1}^{E}}{M_{E}^{2}}\partial^{2}G_{E}F_{\mu\nu}F^{\mu\nu}+\frac{\mathcal{C}_{1}^{E}}{M_{E}^{2}}\partial^{\mu}\partial^{\rho}G_{E}F_{\mu\nu}F_{\rho}^{\ \nu}+\mathcal{D}_{1}^{E}G_{E}F_{\mu\nu}F^{\mu\nu}\right),\\
S_{1}^{HL-G_{E}}= & 2\mathrm{Tr}\sum_{m,n}\int d^{4}xdZ\phi_{m}\phi_{n}\bigg(\frac{\mathcal{B}_{1}^{E}}{M_{E}^{2}}\partial^{2}G_{E}\partial_{[\mu}Q_{\nu],m}^{\dagger}\partial^{[\mu}Q_{\ ,n}^{\nu]}\nonumber \\
 & +\frac{\mathcal{C}_{1}^{E}}{M_{E}^{2}}\eta^{\nu\sigma}\partial^{\mu}\partial^{\rho}G_{E}\partial_{[\mu}Q_{\nu],m}^{\dagger}\partial_{[\rho}Q_{\sigma],n}+\mathcal{D}_{1}^{E}G_{E}\partial_{[\mu}Q_{\nu],m}^{\dagger}\partial^{[\mu}Q_{\ ,n}^{\nu]}\bigg),\\
S_{1}^{HL-L-G_{E}}= & \mathrm{Tr}\sum_{m,n}\int d^{4}xdZ\bigg(\frac{\mathcal{B}_{1}^{E}}{M_{E}^{2}}\eta^{\mu\rho}\eta^{\nu\sigma}\partial^{2}G_{E}\mathcal{I}_{\mu\nu\rho\sigma}+\frac{\mathcal{C}_{1}^{E}}{M_{E}^{2}}\eta^{\nu\sigma}\partial^{\mu}\partial^{\rho}G_{E}\mathcal{I}_{\mu\nu\rho\sigma}\nonumber \\
 & +\mathcal{D}_{1}^{E}\eta^{\mu\rho}\eta^{\nu\sigma}G_{E}\mathcal{I}_{\mu\nu\rho\sigma}\bigg),
\end{align}
where

\begin{align}
\mathcal{I}_{\mu\nu\rho\sigma}= & \sum_{m,n}\bigg(-Q_{[\mu,m}Q_{\nu],n}^{\dagger}F_{\rho\sigma}-F_{\mu\nu}Q_{[\rho,m}Q_{\sigma],n}^{\dagger}+2\partial_{[\mu}Q_{\nu],m}^{\dagger}A_{[\rho}Q_{\sigma],n}+2Q_{[\mu,m}^{\dagger}A_{\nu]}\partial_{[\rho}Q_{\sigma],n}\nonumber \\
 & +Q_{[\mu,m}^{\dagger}A_{\nu]}A_{[\rho}Q_{\sigma],n}\bigg)\phi_{m}\phi_{n}.
\end{align}
Note that $\mathcal{I}_{\mu\nu\rho\sigma}$ satisfies the relation
$\mathcal{I}_{\mu\nu\rho\sigma}=-\mathcal{I}_{\mu\nu\sigma\rho}=-\mathcal{I}_{\nu\mu\rho\sigma},\ \mathrm{Tr}\mathcal{I}_{\mu\nu\rho\sigma}=\mathrm{Tr}\mathcal{I}_{\rho\sigma\mu\nu}$.
And we also have,

\begin{align}
S_{2}^{HL-L}= & 2M_{KK}^{2}\mathrm{Tr}\sum_{m,n}\int d^{4}xdZ\mathcal{A}_{2}\eta^{\mu\nu}\mathcal{II}_{\mu\nu},\\
S_{2}^{L-G_{E}}= & M_{KK}^{2}\mathrm{Tr}\int d^{4}xdZ\left[\frac{\mathcal{B}_{2}^{E}}{M_{E}^{2}}\partial^{\mu}\partial^{\nu}G_{E}F_{\mu Z}F_{\nu Z}+\left(\frac{\mathcal{C}_{2}^{E}}{M_{E}^{2}}\partial^{2}G_{E}+\mathcal{D}_{2}^{E}G_{E}\right)F_{\mu Z}F_{\ Z}^{\mu}\right],\\
S_{2}^{HL-G_{E}}= & 2M_{KK}^{2}\mathrm{Tr}\sum_{m,n}\int d^{4}xdZ\bigg[\left(\frac{\mathcal{C}_{2}^{E}}{M_{E}^{2}}\partial^{2}G_{E}\eta^{\mu\nu}+\frac{\mathcal{B}_{2}^{E}}{M_{E}^{2}}\partial^{\mu}\partial^{\nu}G_{E}+\mathcal{D}_{2}^{E}G_{E}\eta^{\mu\nu}\right)\nonumber \\
 & \times\left(U_{KK}^{2}\partial_{\mu}Q_{m}^{\dagger}\partial_{\nu}Q_{n}\tilde{\phi}_{m}\tilde{\phi}_{n}+Q_{\mu,m}^{\dagger}Q_{\nu,m}\phi_{m}^{\prime}\phi_{n}^{\prime}\right)\bigg],\\
S_{2}^{HL-L-G_{E}}= & M_{KK}^{2}\mathrm{Tr}\sum_{m,n}\int d^{4}xdZ\left[\frac{\mathcal{B}_{2}^{E}}{M_{E}^{2}}\partial^{\mu}\partial^{\nu}G_{E}\mathcal{II}_{\mu\nu}+\frac{\mathcal{C}_{2}^{E}}{M_{E}^{2}}\eta^{\mu\nu}\partial^{2}G_{E}\mathcal{II}_{\mu\nu}+\mathcal{D}_{2}^{E}G_{E}\eta^{\mu\nu}\mathcal{II}_{\mu\nu}\right],
\end{align}
where

\begin{align}
\mathcal{II}_{\mu\nu}= & \sum_{m,n}\bigg[-U_{KK}\phi_{m}\tilde{\phi}_{n}Q_{\mu,m}Q_{n}^{\dagger}F_{\nu Z}+U_{KK}\phi_{m}\tilde{\phi}_{n}Q_{n}Q_{\mu,m}^{\dagger}F_{\nu Z}-U_{KK}\phi_{m}\tilde{\phi}_{n}F_{\mu Z}Q_{\nu,m}Q_{n}^{\dagger}\nonumber \\
 & +U_{KK}\phi_{m}\tilde{\phi}_{n}F_{\mu Z}Q_{n}Q_{\nu,m}^{\dagger}+2\bigg(U_{KK}Q_{\mu,m}^{\dagger}A_{Z}\partial_{\nu}Q_{n}\phi_{m}\tilde{\phi}_{n}-U_{KK}^{2}Q_{m}^{\dagger}A_{\mu}\partial_{\nu}Q_{n}\tilde{\phi}_{m}\tilde{\phi}_{n}\nonumber \\
 & -Q_{\mu,m}^{\dagger}A_{Z}Q_{\nu,n}\phi_{m}\phi_{n}^{\prime}+U_{KK}Q_{m}^{\dagger}A_{\mu}Q_{\nu,n}\tilde{\phi}_{m}\phi_{n}^{\prime}+U_{KK}^{2}\partial_{\mu}Q_{m}^{\dagger}A_{\nu}Q_{n}\tilde{\phi}_{m}\tilde{\phi}_{n}\nonumber \\
 & -U_{KK}Q_{\mu,m}^{\dagger}A_{\nu}Q_{n}\phi_{m}^{\prime}\tilde{\phi}_{n}-U_{KK}\partial_{\mu}Q_{m}^{\dagger}A_{Z}Q_{\nu,n}\tilde{\phi}_{m}\phi_{n}+Q_{\mu,m}^{\dagger}A_{Z}Q_{\nu,n}\phi_{m}^{\prime}\phi_{n}\nonumber \\
 & +U_{KK}Q_{\mu,m}^{\dagger}A_{Z}A_{\nu}Q_{n}\phi_{m}\tilde{\phi}_{n}-U_{KK}^{2}Q_{m}^{\dagger}\tilde{\phi}_{m}A_{\mu}A_{\nu}Q_{n}\tilde{\phi}_{n}-Q_{\mu,m}^{\dagger}A_{Z}^{2}Q_{\nu,n}\phi_{m}\phi_{n}\nonumber \\
 & +U_{KK}Q_{m}^{\dagger}A_{\mu}A_{Z}Q_{\nu,n}\tilde{\phi}_{m}\phi_{n}\bigg)\bigg].
\end{align}
For the terms in $S_{3}$, respectively they are,

\begin{align}
S_{3}^{L-G_{E}}= & \frac{M_{KK}^{2}}{M_{E}^{2}}\mathrm{Tr}\int d^{4}xdZc\left(Z\right)\partial_{\sigma}G_{E}F^{\rho\sigma}F_{\rho Z},\\
S_{3}^{HL-G_{E}}= & \frac{2M_{KK}^{2}}{M_{E}^{2}}\mathrm{Tr}\sum_{m,n}\int d^{4}xdZc\left(Z\right)\partial_{\sigma}G_{E}\left(\phi_{m}\tilde{\phi}_{n}U_{KK}\partial^{[\rho}Q_{\ ,m}^{\sigma]\dagger}\partial_{\rho}Q_{n}-\phi_{m}\phi_{n}^{\prime}\partial^{[\rho}Q_{\ ,m}^{\sigma]\dagger}Q_{\rho,n}\right),\\
S_{3}^{HL-L-G_{E}}= & \frac{M_{KK}^{2}}{M_{E}^{2}}\mathrm{Tr}\sum_{m,n}\int d^{4}xdZc\left(Z\right)\partial_{\sigma}G_{E}\bigg[-\phi_{m}\phi_{n}Q_{\ ,m}^{[\rho}Q_{\ ,n}^{\sigma]\dagger}F_{\rho Z}-F^{\rho\sigma}Q_{\rho,m}Q_{n}^{\dagger}U_{KK}\phi_{m}\tilde{\phi}_{n}\nonumber \\
 & +F^{\rho\sigma}Q_{n}Q_{\rho,m}^{\dagger}U_{KK}\phi_{m}\tilde{\phi}_{n}+2\bigg(U_{KK}\partial^{[\rho}Q_{\ ,m}^{\sigma]\dagger}\partial_{\rho}Q_{n}\phi_{m}\tilde{\phi}_{n}-\partial^{[\rho}Q_{\ ,m}^{\sigma]\dagger}Q_{\rho,n}\phi_{m}\phi_{n}^{\prime}\nonumber \\
 & +U_{KK}Q_{\ ,m}^{[\rho\dagger}A^{\sigma]}\partial_{\rho}Q_{n}\phi_{m}\tilde{\phi}_{n}-Q_{\ ,m}^{[\rho\dagger}A^{\sigma]}Q_{\rho,n}\phi_{m}\phi_{n}^{\prime}+U_{KK}\partial^{[\rho}Q_{\ ,m}^{\sigma]\dagger}A_{\rho}Q_{n}\phi_{m}\tilde{\phi}_{n}\nonumber \\
 & -\partial^{[\rho}Q_{\ ,m}^{\sigma]\dagger}A_{Z}Q_{\rho,n}\phi_{m}\phi_{n}+U_{KK}Q_{\ ,m}^{[\rho\dagger}A^{\sigma]}A_{\rho}Q_{n}\phi_{m}\tilde{\phi}_{n}-Q_{\ ,m}^{[\rho\dagger}A^{\sigma]}A_{Z}Q_{\rho,n}\phi_{m}\phi_{n}\bigg)\bigg],\label{eq:56}
\end{align}
and all the coefficients are given as,

\begin{align}
 & \mathcal{A}_{1}=-\frac{\kappa}{2}K{}^{-1/3},\ \mathcal{B}_{1}^{E}=\frac{1}{4}\kappa\tilde{H}_{E}K{}^{-1/3},\ \mathcal{C}_{1}^{E}=-\kappa\tilde{H}_{E}K{}^{-1/3},\ \mathcal{D}_{1}^{E}=-\frac{1}{16}\frac{5K-14}{5K-2}\kappa H_{E}K{}^{-1/3},\nonumber \\
 & \mathcal{A}_{2}=-\kappa K,\ \mathcal{B}_{2}^{E}=-\kappa K\tilde{H}_{E},\ \mathcal{C}_{2}^{E}=\frac{1}{2}\kappa K\tilde{H}_{E},\ \mathcal{D}_{2}^{E}=-\frac{3}{8}\frac{5K+2}{5K-2}\kappa KH_{E},\nonumber \\
 & c\left(Z\right)=60\kappa\frac{ZK}{\left(5K-2\right)^{2}}H_{E},\ \bar{H}_{E}=\left(\frac{1}{4}+\frac{3}{5K-2}\right)H_{E}.
\end{align}
Additionally we have another coupling from the mass terms (\ref{eq:34})
which is,

\begin{align}
S_{m}^{HL-G_{E}}=- & \frac{v_{V}^{2}}{M_{E}^{2}}\mathrm{Tr}\sum_{m.n}\int d^{4}x\left[s_{mn}^{\left(1\right)}\partial^{2}G_{E}Q_{\mu,m}^{\dagger}Q_{\ ,n}^{\mu}+s_{mn}^{\left(2\right)}\partial^{\mu}\partial^{\nu}G_{E}Q_{\mu,m}^{\dagger}Q_{\nu,n}+s_{mn}^{\left(3\right)}M_{E}^{2}G_{E}Q_{\mu,m}^{\dagger}Q_{\ ,n}^{\mu}\right]\nonumber \\
 & +\frac{v_{S}^{2}}{M_{E}^{2}}\mathrm{Tr}\sum_{m.n}\int d^{4}x\left[s_{mn}^{\left(4\right)}\partial^{2}G_{E}Q_{m}^{\dagger}Q_{n}+s_{mn}^{\left(5\right)}M_{E}^{2}G_{E}Q_{m}^{\dagger}Q_{n}\right]\nonumber \\
 & +\alpha_{E}^{2}\mathrm{Tr}\sum_{m.n}\int d^{4}xs_{mn}^{\left(6\right)}M_{KK}\partial^{\mu}G_{E}\left(Q_{\mu,m}^{\dagger}Q_{n}+Q_{n}^{\dagger}Q_{\mu,m}\right),\label{eq:58}
\end{align}
where $\alpha_{E}=\frac{\left(1+N_{f}\right)v}{N_{f}M_{E}}\frac{v_{S}}{v_{V}}$
and the coefficients are,

\begin{align}
 & s_{mn}^{\left(1\right)}=\frac{\kappa}{2}\int dZK^{1/6}\tilde{H}_{E}\phi_{m}\phi_{n},\ s_{mn}^{\left(2\right)}=-\kappa\int dZK^{1/6}\tilde{H}_{E}\phi_{m}\phi_{n},\nonumber \\
 & s_{mn}^{\left(3\right)}=-\kappa\int dZ\frac{5Z^{2}H_{E}}{2\left(5K-2\right)}K^{1/6}\phi_{m}\phi_{n},\ s_{mn}^{\left(4\right)}=-\frac{\kappa}{2}\int dZK^{3/2}\tilde{H}_{E}\tilde{\phi}_{m}\tilde{\phi}_{n}\nonumber \\
 & s_{mn}^{\left(5\right)}=\kappa\int dZ\frac{15K^{5/2}H_{E}}{4\left(5k-2\right)}\tilde{\phi}_{m}\tilde{\phi}_{n},\ s_{mn}^{\left(6\right)}=-\kappa\int dZ\frac{10ZK^{3/2}}{\left(5K-2\right)^{2}}H_{E}\phi_{m}\tilde{\phi}_{n}.
\end{align}
The total action of the exotic scalar glueball and various mesons
could be obtained by collecting all the terms in (\ref{eq:40}) and
(\ref{eq:58}) with the kinetic term of the glueball in (\ref{eq:12}).

\subsection{Involving the dilatonic scalar glueball }

The effective action of the dilatonic glueball and mesons could be
directly obtained by plugging (\ref{eq:42}) (\ref{eq:43}) with the
dilatonic polarizations (\ref{eq:14}) into action (\ref{eq:41}).
We find that the interaction terms $S_{i}^{HL-L}$ in (\ref{eq:41})
remains since they are independent on the metric fluctuations. So
the interaction terms of dilatonic glueball and mesons can be collected
as

\begin{equation}
S_{\mathrm{int}}^{G_{D}}=\sum_{i=1}^{2}\left(S_{i}^{L-G_{D}}+S_{i}^{HL-G_{D}}+S_{i}^{HL-L-G_{D}}\right)+S_{m}^{HL-G_{D}},
\end{equation}
where $S_{i}^{L-G_{D}},S_{i}^{HL-G_{D}},S_{i}^{HL-L-G_{D}}$ takes
the same formulas as in the case of the exotic scalar glueball while
the coefficients have to be replaced by $\mathcal{B}_{1,2}^{E},\mathcal{C}_{1,2}^{E},\mathcal{D}_{1,2}^{E}\rightarrow\mathcal{B}_{1,2}^{D},\mathcal{C}_{1,2}^{D},\mathcal{D}_{1,2}^{D}$.
And the corresponding coefficients are given as,

\begin{align}
 & \mathcal{B}_{1}^{D}=\frac{1}{4}\kappa H_{D}K{}^{-1/3},\ \mathcal{C}_{1}^{D}=-\kappa H_{D}K{}^{-1/3},\ \mathcal{D}_{1}^{D}=\frac{3}{4}\kappa H_{D}K{}^{-1/3}\nonumber \\
 & \mathcal{B}_{2}^{D}=-\kappa KH_{D},\ \mathcal{C}_{2}^{D}=\frac{1}{2}\kappa KH_{D},\ \mathcal{D}_{2}^{D}=\mathcal{C}_{2}^{D}.\label{eq:61}
\end{align}
The coupling term $S_{m}^{HL-G_{D}}$ for the dilatonic glueball from
the mass terms (\ref{eq:34}) also takes the same formula as in (\ref{eq:58})
while the coefficients need to be replaced as $s_{mn}^{\left(1,2...6\right)}\rightarrow d_{mn}^{\left(1,2...6\right)}$
and $d_{mn}^{\left(1,2...6\right)}$'s are given as,

\begin{align}
 & d_{mn}^{\left(1\right)}=\frac{\kappa}{2}\int dZK^{1/6}H_{D}\phi_{m}\phi_{n},\ d_{mn}^{\left(2\right)}=-\kappa\int dZK^{1/6}H_{D}\phi_{m}\phi_{n},\nonumber \\
 & d_{mn}^{\left(3\right)}=2\kappa\int dZK^{1/6}H_{D}\phi_{m}\phi_{n},\ d_{mn}^{\left(4\right)}=\frac{\kappa}{2}\int dZK^{3/2}H_{D}\tilde{\phi}_{m}\tilde{\phi}_{n}\nonumber \\
 & d_{mn}^{\left(5\right)}=\kappa\int dZK^{3/2}H_{D}\tilde{\phi}_{m}\tilde{\phi}_{n},\ d_{mn}^{\left(6\right)}=0.\label{eq:62}
\end{align}
So the total action for the interaction of dilatonic scalar glueball
and meson could also be collected from (\ref{eq:44}) - (\ref{eq:56})
and (\ref{eq:58}) while the coefficients are replaced by (\ref{eq:61})
(\ref{eq:62}).

\subsection{Involving the tensor glueball}

The interaction terms of tensor glueball and mesons can be collected
by inserting (\ref{eq:42}) (\ref{eq:43}) with the polarizations
(\ref{eq:15}) into (\ref{eq:29}) (\ref{eq:34}). As before, let
us rewrite the interaction terms involving the tensor glueball in
the action as,

\begin{align}
S_{\mathrm{int}}^{T} & =\sum_{i=1}^{2}\left(S_{i}^{L-T}+S_{i}^{H-T}+S_{i}^{HL-L-T}\right)+S_{m}^{HL-T},
\end{align}
where their explicit formulas are collected as,

\begin{align}
S_{1}^{L-T}= & \mathrm{Tr}\int d^{4}xdZB_{1}^{T}T^{\mu\rho}\eta^{\nu\sigma}F_{\mu\nu}F_{\rho\sigma},\\
S_{1}^{HL-T}= & 2\mathrm{Tr}\sum_{m,n}\int d^{4}xdZB_{1}^{T}\phi_{m}\phi_{n}T^{\nu\sigma}\eta^{\mu\rho}\partial_{[\mu}Q_{\nu,m]}^{\dagger}\partial_{[\rho}Q_{\sigma,n]},\\
S_{1}^{HL-L-T}= & \mathrm{Tr}\sum_{m,n}\int d^{4}xdZB_{1}^{T}T^{\nu\sigma}\eta^{\mu\rho}\mathcal{I}_{\mu\nu\rho\sigma},
\end{align}
and

\begin{align}
S_{2}^{L-T}= & M_{KK}^{2}\mathrm{Tr}\int d^{4}xdZ\mathcal{B}_{2}^{T}T^{\mu\nu}F_{\mu Z}F_{\nu Z},\\
S_{2}^{HL-T}= & 2M_{KK}^{2}\mathrm{Tr}\sum_{m,n}\int d^{4}xdZ\mathcal{B}_{2}^{T}T^{\mu\nu}\left(U_{KK}^{2}\partial_{\mu}Q_{m}^{\dagger}\partial_{\nu}Q_{n}\tilde{\phi}_{m}\tilde{\phi}_{n}+Q_{\mu,m}^{\dagger}Q_{\nu,m}\phi_{m}^{\prime}\phi_{n}^{\prime}\right),\\
S_{2}^{HL-L-T}= & M_{KK}^{2}\mathrm{Tr}\int d^{4}xdZ\mathcal{B}_{2}^{T}T^{\mu\nu}\mathcal{II}_{\mu\nu}.
\end{align}
The coupling from the mass term (\ref{eq:34}) is given as, 

\begin{align}
S_{m}^{HL-T}=- & \sum_{m.n}v_{V}^{2}t_{mn}\mathrm{Tr}\int d^{4}xT^{\mu\nu}Q_{\mu,m}^{\dagger}Q_{\nu,n},
\end{align}
where the corresponding coefficients in above actions are,

\begin{equation}
\mathcal{B}_{1}^{T}=-\kappa K^{-1/3}H_{T},\ \mathcal{B}_{2}^{T}=-\kappa KH_{T},\ t_{mn}=-\kappa\int dZK^{1/6}H_{D}\phi_{m}\phi_{n}.
\end{equation}

\section{Interaction of glueball and scalar $D^{0}$ meson in holography}

Since the lowest heavy-light meson is recognized as D-meson, let us
identify the lowest heavy-light meson field in our theory as the lowest
D-meson field (i.e. the $D^{0}$ meson) and investigate the decay
of glueball to them holographically in this section.

\subsection{The mode eigenfunctions and the choice of parameters}

The eigenfunctions $\left\{ \phi_{m},\tilde{\phi}_{n}\right\} $ (\ref{eq:36})
have to satisfy the normalization condition (\ref{eq:37}) which is
exactly the same as the normalization of $\left\{ \psi_{n}\right\} $
for $n\geq1$ presented in \cite{key-12},

\begin{align}
2\kappa\int_{-\infty}^{+\infty}dZK\left(Z\right)^{-1/3}\psi_{m}\left(Z\right)\psi_{n}\left(Z\right)= & \delta_{mn},\nonumber \\
2\kappa\int_{-\infty}^{+\infty}dZK\left(Z\right)\partial_{Z}\psi_{m}\partial_{Z}\psi_{n}= & \lambda_{n}\delta_{mn}.
\end{align}
From these, the functions $\left\{ \psi_{n}\right\} $ for $n\geq1$
could be generated by the following eigen equation,

\begin{equation}
-K^{1/3}\partial_{Z}\left(K\partial_{Z}\psi_{n}\right)=\lambda_{n}\psi_{n},\ \psi_{n}\left(Z\rightarrow\pm\infty\right)=0,\label{eq:73}
\end{equation}
Thus it is seen that we can choose $\phi_{m}=\frac{\psi_{m}}{\sqrt{2}},\tilde{\phi}_{n}=\frac{1}{\sqrt{2}}\lambda_{n}^{-1/2}M_{KK}^{-1}U_{KK}^{-1}\partial_{Z}\psi_{n}$
for all $m,n\geq1$. Note that there still exists a function $\tilde{\phi}_{0}$
orthogonal to $\left\{ \tilde{\phi}_{n}\right\} $ for all $n\geq1$
which is given as,

\begin{equation}
\tilde{\phi}_{0}=\frac{1}{\sqrt{2}}\psi_{0}^{\prime}\left(Z\right)=\frac{1}{2\sqrt{\pi\kappa}}\frac{1}{U_{KK}M_{KK}}\frac{1}{K\left(Z\right)},
\end{equation}
while $\psi_{0}$ does not satisfy the eigen equation (\ref{eq:73}).
So the complete set of eigenfunctions $\left\{ \phi_{m}\right\} $
can be obtained by re-defining $\phi_{m}=\frac{\psi_{m+1}}{\sqrt{2}}$
for all $m\geq0$ thus $\Lambda_{0}=1,\Lambda_{m}=\lambda_{m},m>0$.
Hence we are going to continue the discussion with this definition.

In order to match the experimental value of the $\rho$ meson mass
$m_{\rho}=\sqrt{\lambda_{1}}M_{KK}\simeq776\mathrm{MeV}$ where $\lambda_{1}$
is numerically evaluated as $\lambda_{1}=0.669314...$, we first need
to fix the Kaluza-Klein mass to $M_{KK}=949\mathrm{MeV}$. Then let
us identify the lowest mode of the heavy-light meson field $Q_{\mu,n=0},Q_{n=0}$
as the vector meson $D_{\mu}^{0}$ and the pseudoscalar meson $D^{0}$
whose mass is given as $M\left(D_{\mu}^{0}\right)\simeq2007\mathrm{MeV},\ M\left(D^{0}\right)\simeq1865\mathrm{MeV}$
by experiments. In this sense, we further consider the case of $N_{f}=2$
for D-meson, so the parameters in (\ref{eq:38}) (\ref{eq:39}) can
be numerically fixed as

\begin{equation}
M_{S}\simeq1754\mathrm{MeV},\ M_{V}\simeq2007\mathrm{MeV},\ M_{KK}R\simeq1.048,\ v\simeq35195.4\lambda^{-1}N_{c}^{-1}M_{KK}.
\end{equation}
On the other hand, the glueball mass in holography is determined by
solving the eigen equations (\ref{eq:11}) (\ref{eq:16}) for exotic,
dilatonic and tensor glueball respectively. For the reader's convenience,
we collect the mass spectrum of various glueballs in Table \ref{tab:2}.
It is clear that the glueball decay occurs when the mass relation
$m_{\mathrm{glueball}}/m_{\mathrm{D-meson}}\simeq2$ is satisfied
because the effective holographic action we discussed is quadratic
of the heavy-light meson fields. Accordingly we will choose the excited
mass of $n=3$ as $M_{E}=\sqrt{154.963/9}M_{KK}\simeq3938\mathrm{MeV},\ M_{D}=M_{T}=\sqrt{162.699/9}M_{KK}\simeq4035\mathrm{MeV}$
for the exotic, dilatonic and tensor glueball respectively. With these
values for the parameters, the boundary condition for the eigen equations
(\ref{eq:11}) (\ref{eq:16}) is numerically evaluated as,

\begin{equation}
H_{E}\left(r_{KK}\right)^{-1}\simeq0.00692402\lambda^{1/2}N_{c}M_{KK},\ \ H_{D,T}\left(r_{KK}\right)^{-1}\simeq0.0211777\lambda^{1/2}N_{c}M_{KK}.
\end{equation}
And we will use the parameters above to investigate the decay of glueball
the lowest D-meson.

\begin{table}
\begin{centering}
\begin{tabular}{|c|c|c|c|c|c|c|}
\hline 
Mode & $\mathrm{S}_{4}$ & $\mathrm{T}_{4}$ & $\mathrm{V}_{4}$ & $\mathrm{N}_{4}$ & $\mathrm{M}_{4}$ & $\mathrm{L}_{4}$\tabularnewline
$J^{PC}$ & $0^{++}$ & $0^{++}/2^{++}$ & $0^{-+}$ & $1^{+-}$ & $1^{--}$ & $0^{++}$\tabularnewline
\hline 
$n=0$ & 7.30835 & 22.0966 & 31.9853 & 53.3758 & 83.0449 & 115.002\tabularnewline
\hline 
$n=1$ & 46.9855 & 55.5833 & 72.4793 & 109.446 & 143.581 & 189.632\tabularnewline
\hline 
$n=2$ & 94.4816 & 102.452 & 126.144 & 177.231 & 217.397 & 227.283\tabularnewline
\hline 
$n=3$ & 154.963 & 162.699 & 193.133 & 257.959 & 304.531 & 378.099\tabularnewline
\hline 
$n=4$ & 228.709 & 236.328 & 273.482 & 351.895 & 405.011 & 492.171\tabularnewline
\hline 
\end{tabular}
\par\end{centering}
\caption{\label{tab:2}Glueball mass spectrum $m_{n}^{2}$ of $\mathrm{AdS}_{7}$
in the units of $r_{KK}^{2}/L^{4}=M_{KK}^{2}/9$.}
\end{table}

\subsection{Glueball decay to the lowest scalar $D^{0}\bar{D}^{0}$ meson}

In this section, let us compute the decay width of the glueball to
the lowest heavy-light mesons i.e. the (pseudo) scalar meson ($D^{0}\bar{D}^{0}$).
The expression of the decay width for the scalar glueball into $D\bar{D}$
is given by\footnote{We have employed the same convention as those in \cite{key-26}.},

\begin{equation}
\Gamma_{G_{E,D}\rightarrow D\bar{D}}=\frac{3\left|\mathrm{\mathbf{p}}\right|}{8\pi M_{E,D}^{2}}\left|\mathcal{M}_{E,D}\right|^{2},\label{eq:77}
\end{equation}
where $\left|\boldsymbol{\mathrm{p}}\right|=M_{E,D}/2$ represents
the 3-momentum of one $D^{0}$ in the rest frame of the glueball.
For the tensor glueball we however need the average over its polarizations.
Hence let us integrate over the orientation of the coordinates by
choosing a fixed polarization $\epsilon^{11}=-\epsilon^{22}=1$. So
it leads to the decay width as.

\begin{equation}
\Gamma_{G_{T}\rightarrow D\bar{D}}=\frac{3\left|\mathrm{\mathbf{p}}\right|}{8\pi M_{T}^{2}}\int\frac{d\Omega}{4\pi}\left|\mathcal{M}_{T}\right|^{2},\ \left|\mathrm{\mathbf{p}}\right|=\frac{M_{T}}{2}.
\end{equation}
Note that ($D^{0}\bar{D}^{0}$) will be denoted as ($D\bar{D}$) for
simplicity in the following discussion of this section.

\subsubsection*{A. exotic scalar glueball }

The relevant part of the action involving the exotic glueball and
$D^{0}\bar{D}^{0}$ can be collected from (\ref{eq:44}) - (\ref{eq:58})
which is,

\begin{align}
S^{G_{E}\rightarrow D\bar{D}}= & \mathrm{Tr}\int d^{4}x\bigg[c_{2}^{E}\partial^{2}G_{E}\partial_{\mu}\bar{D}\partial^{\mu}D+b_{2}^{E}\partial^{\mu}\partial^{\nu}G_{E}\partial_{\mu}\bar{D}\partial_{\nu}D+d_{2}^{E}G_{E}\partial_{\mu}\bar{D}\partial^{\mu}D\nonumber \\
 & +\frac{v_{S}^{2}}{M_{E}^{2}}s_{00}^{\left(4\right)}\partial^{2}G_{E}\bar{D}D+v_{S}^{2}s_{00}^{\left(5\right)}G_{E}\bar{D}D\bigg],\label{eq:78}
\end{align}
where the coupling constants could be numerically evaluated as,

\begin{align}
c_{2}^{E} & =\frac{2M_{KK}^{2}U_{KK}^{2}}{M_{E}^{2}}\int dZ\mathcal{C}_{2}^{E}\tilde{\phi}_{0}\tilde{\phi}_{0}\simeq-0.403\lambda^{-1/2}N_{c}^{-1}M_{KK}^{-3},\nonumber \\
b_{2}^{E} & =\frac{2M_{KK}^{2}U_{KK}^{2}}{M_{E}^{2}}\int dZ\mathcal{B}_{2}^{E}\tilde{\phi}_{0}\tilde{\phi}_{0}\simeq0.806\lambda^{-1/2}N_{c}^{-1}M_{KK}^{-3},\nonumber \\
d_{2}^{E} & =2M_{KK}^{2}U_{KK}^{2}\int dZ\mathcal{D}_{2}^{E}\tilde{\phi}_{0}\tilde{\phi}_{0}\simeq9.432\lambda^{-1/2}N_{c}^{-1}M_{KK}^{-1},\nonumber \\
s_{00}^{(4)} & \simeq-12.020\lambda^{-1/2}N_{c}^{-1}M_{KK}^{-1},\ \ s_{00}^{\left(5\right)}\simeq-21.749\lambda^{-1/2}N_{c}^{-1}M_{KK}^{-1}.
\end{align}
By omitting the terms which vanishes when $G_{E}$ is onshell in (\ref{eq:78}),
the amplitude is calculated as,

\begin{align}
\left|\mathcal{M}_{E}\right| & =\left|\left[2b_{2}^{E}M_{E}^{2}+2\left(c_{2}^{E}M_{E}^{2}+d_{2}^{E}\right)\right]p_{0}q_{0}-2\left(c_{2}^{E}M_{E}^{2}+d_{2}^{E}\right)\mathbf{p}\cdot\mathbf{q}+v_{S}^{2}\left[s_{00}^{\left(4\right)}+s_{00}^{\left(5\right)}\right]\right|\nonumber \\
 & =\left|\frac{M_{E}^{2}}{4}\left[2b_{2}^{E}M_{E}^{2}+4\left(c_{2}^{E}M_{E}^{2}+d_{2}^{E}\right)\right]+v_{S}^{2}\left[s_{00}^{\left(4\right)}+s_{00}^{\left(5\right)}\right]\right|,
\end{align}
and the decay rate is accordingly obtained as,

\begin{equation}
\Gamma_{G_{E}\rightarrow D\bar{D}}\big/M_{E}=\frac{3\left|\mathrm{\mathbf{p}}\right|}{8\pi M_{E}^{3}}\left|\mathcal{M}_{E}\right|^{2}\simeq\frac{5.310}{\lambda N_{c}^{2}}\frac{M_{E}^{2}}{M_{KK}^{2}}+\mathcal{O}\left(N_{c}^{-4}\right)\simeq\frac{91.421}{\lambda N_{c}^{2}}.
\end{equation}

\subsubsection*{B. dilatonic scalar glueball}

For the decay of the dilatonic glueball, we have the same formula
for the action collected from (\ref{eq:44}) - (\ref{eq:58}),

\begin{align}
S^{G_{D}\rightarrow D\bar{D}}= & \mathrm{Tr}\int d^{4}x\bigg[c_{2}^{D}\partial^{2}G_{D}\partial_{\mu}\bar{D}\partial^{\mu}D+b_{2}^{D}\partial^{\mu}\partial^{\nu}G_{D}\partial_{\mu}\bar{D}\partial_{\nu}D+d_{2}^{D}G_{D}\partial_{\mu}\bar{D}\partial^{\mu}D\nonumber \\
 & +\frac{v_{S}^{2}}{M_{E}^{2}}d_{mn}^{\left(4\right)}\partial^{2}G_{D}\bar{D}D+v_{S}^{2}d_{mn}^{\left(5\right)}G_{D}\bar{D}D\bigg],
\end{align}
while the coupling constants are numerically evaluated as,

\begin{align}
c_{2}^{D} & =\frac{2M_{KK}^{2}U_{KK}^{2}}{M_{D}^{2}}\int dZ\mathcal{C}_{2}^{D}\tilde{\phi}_{0}\tilde{\phi}_{0}\simeq0.105\lambda^{-1/2}N_{c}^{-1}M_{KK}^{-3},\nonumber \\
b_{2}^{D} & =\frac{2M_{KK}^{2}U_{KK}^{2}}{M_{D}^{2}}\int dZ\mathcal{B}_{2}^{D}\tilde{\phi}_{0}\tilde{\phi}_{0}\simeq-0.211\lambda^{-1/2}N_{c}^{-1}M_{KK}^{-3},\nonumber \\
d_{2}^{D} & =2M_{KK}^{2}U_{KK}^{2}\int dZ\mathcal{D}_{2}^{D}\tilde{\phi}_{0}\tilde{\phi}_{0}\simeq1.907\lambda^{-1/2}N_{c}^{-1}M_{KK}^{-1},\nonumber \\
d_{00}^{(4)} & \simeq7.859\lambda^{-1/2}N_{c}^{-1}M_{KK}^{-1},\ \ d_{00}^{\left(5\right)}\simeq6.454\lambda^{-1/2}N_{c}^{-1}M_{KK}^{-1}.
\end{align}
Omitting the terms when $G_{D}$ is onshell in (\ref{eq:78}), the
amplitude is calculated as,

\begin{align}
\left|\mathcal{M}_{D}\right| & =\left|\left[2b_{2}^{D}M_{D}^{2}+2\left(c_{2}^{D}M_{D}^{2}+d_{2}^{D}\right)\right]p_{0}q_{0}-2\left(c_{2}^{D}M_{D}^{2}+d_{2}^{D}\right)\mathbf{p}\cdot\mathbf{q}+v_{S}^{2}\left[d_{00}^{\left(4\right)}+d_{00}^{\left(5\right)}\right]\right|\nonumber \\
 & =\left|\frac{M_{D}^{2}}{4}\left[2b_{2}^{D}M_{D}^{2}+4\left(c_{2}^{D}M_{D}^{2}+d_{2}^{D}\right)\right]+v_{S}^{2}\left[d_{00}^{\left(4\right)}+d_{00}^{\left(5\right)}\right]\right|,
\end{align}
which leads to the decay rate as,

\begin{equation}
\Gamma_{G_{D}\rightarrow D\bar{D}}\big/M_{D}=\frac{3\left|\mathrm{\mathbf{p}}\right|}{8\pi M_{D}^{3}}\left|\mathcal{M}_{D}\right|^{2}\simeq\frac{0.215}{\lambda N_{c}^{2}}\frac{M_{D}^{2}}{M_{KK}^{2}}+\mathcal{O}\left(N_{c}^{-4}\right)\simeq\frac{3.887}{\lambda N_{c}^{2}}.
\end{equation}

\subsubsection*{C. tensor glueball}

The action involving the tensor glueball and $D\bar{D}$ is very simple,

\begin{equation}
S_{2}^{HL-G_{T}}=b_{2}^{T}\mathrm{Tr}\int d^{4}xT^{\mu\nu}\partial_{\mu}\bar{D}\partial_{\nu}D,
\end{equation}
where the coupling constant is,

\begin{equation}
b_{2}^{T}=2M_{KK}^{2}U_{KK}^{2}\int dZ\mathcal{B}_{2}^{T}\tilde{\phi}_{0}\tilde{\phi}_{0}\simeq-9.343\lambda^{-1/2}N_{c}^{-1}M_{KK}^{-1}.
\end{equation}
Then we could obtain the amplitude when $T_{\mu\nu}$ is onshell as,

\[
\left|\mathcal{M}_{T}\right|=\left|2b_{2}^{T}\left(p_{x}^{2}-p_{y}^{2}\right)\right|,
\]
therefore the decay rate is evaluated as,

\begin{equation}
\Gamma_{T\rightarrow D\bar{D}}\big/M_{T}=\frac{3\left|\mathrm{\mathbf{p}}\right|}{8\pi M_{T}^{3}}\int\frac{d\Omega}{4\pi}\left|\mathcal{M}_{T}\right|^{2}\simeq\frac{0.347}{\lambda N_{c}^{2}}\frac{M_{T}^{2}}{M_{KK}^{2}}\simeq\frac{6.279}{\lambda N_{c}^{2}}.
\end{equation}

\section{Summary and discussion}

By using the Witten-Sakai-Sugimoto model, we study the interaction
of glueball and heavy-light flavoured mesons in holography then extend
the calculations of glueball decay in \cite{key-25,key-26} with heavy
flavour. The model contains only one free dimensionful parameter $M_{KK}$
since it is a top-down approach of string theory. The glueball fields
are identified as the various fluctuations of the bulk geometry produced
by the $N_{c}$ compactified D4-branes and the meson fields are created
by the open strings living in the $N_{f}$ probe $\mathrm{D}8/\overline{\mathrm{D}8}$-branes.
Particularly we construct the D4/D8 configuration as in \cite{key-28,key-29,key-34,key-35}
by introducing an extra pair of probe $\mathrm{D}8/\overline{\mathrm{D}8}$-branes
as ``heavy flavour'' which is separated from the other $N_{f}$
$\mathrm{D}8/\overline{\mathrm{D}8}$-branes with a heavy-light string
stretched between them. In this configuration the multiplets produced
by the heavy-light string could be identified as the heavy-light flavoured
mesons so that the model is able to include their interactions with
glueball.

We numerically evaluate the coupling constants between the various
glueballs (the exotic, dilatonic scalar and tensor glueballs) and
the heavy-light flavoured mesons then compute the associated decay
widths/rates in this model. We find the decay widths of glueballs
are parametrically suppressed by the factor $\lambda^{-1}N_{c}^{-2}$
in the large $N_{c}$ limit while the results vary substantially for
the different decay channels. So it could not give the picture of
``universal narrowness'' although the heavy flavour and large $N_{c}$
limitation are involved in the model. However the decay widths of
the lightest glueball to the lowest light mesons in this model turn
out to be consistent with the experimental data of the $f_{0}\left(1500\right)$
decay \cite{key-25,key-26}, thus our calculation with heavy flavour
could be a further prediction of glueball-meson interaction although
the experimental evidence are currently insufficient\footnote{The $X\left(3020\right)$ \cite{key-38} is the only one glueball
candidate whose energy is close to the glueballs' discussed in our
manuscript. }. Besides, the decay rate of the exotic scalar glueball is turned
out to have a larger width than that of the heavier dilatonic mode
according to our calculation, so our results support partially the
conjecture proposed in \cite{key-26} that is the lowest glueball
model begins with the dilatonic mode and the exotic scalar glueball
should be discarded.

Noteworthily phenomenological models usually require very large gluon
condensates in order to admit only narrow glueball states while the
value of the gluon condensate is quite small in the WSS model. Nevertheless
it could be interesting to introduce the gluon condensates into a
top-down approach of holography as \cite{key-39,key-40,key-41} so
that it would be able to compare with the phenomenological models.

We finally comments that the mixing of glueballs with $q\bar{q}$
meson states would be also interesting to study since the signatures
of glueball content can be strongly obscured by the mixing. But it
might probably require more difficult stringy corrections which are
not captured by the effective action following from the WSS model
in holography. Nonetheless this would be more pivotal to identify
the glueball in the future study.

\section*{Appendix A: The 10d equations of motion in the bulk supergravity}

The WSS model is obtained by the Kaluza-Klein reduction of the 11d
supergravity on a cycle. Since the supersymmetry is broken down in
the model, the bulk geometry of $N_{c}$ D4-branes in the large $N_{c}$
limit is determined by the bosonic part of the type IIA supergravity
whose action is given as,\index{Commands!T!tag@\textbackslash{}tag}

\begin{equation}
S_{\mathrm{IIA}}=\frac{1}{2k_{10}^{2}}\int d^{10}x\sqrt{-g}e^{-2\Phi}\left(\mathcal{R}+4\nabla_{M}\Phi\nabla^{M}\Phi-\frac{1}{2}\left|F_{4}\right|^{2}\right),\tag{A-1}\label{eq:A-1}
\end{equation}
where $2k_{10}^{2}=16G_{10}/g_{s}^{2}=\left(2\pi\right)^{7}l_{s}^{8}$
, $G_{10}$ is 10d Newton constant and $F_{4}=dC_{3}$ is the field
strength of the Romand-Romand (R-R) 3-form $C_{3}$ with,

\begin{equation}
\left|F_{4}\right|^{2}=\frac{1}{4!}F_{ABCD}F^{ABCD}.\tag{A-2}
\end{equation}

The equations of motion could be obtained by varying the action (\ref{eq:A-1})
with respect to the metric $g_{MN}$, the dilaton field $\Phi$ and
the R-R form $C_{3}$ which are,

\begin{align}
\mathcal{R}_{MN}-\frac{1}{2}g_{MN}\mathcal{R}+2\nabla_{M}\Phi\nabla_{N}\Phi-\frac{e^{2\Phi}}{4!}\left(F_{M}^{\ LPQ}F_{NLPQ}-3!g_{MN}\left|F_{4}\right|^{2}\right) & =0,\nonumber \\
\mathcal{R}+4\nabla_{M}\nabla^{M}\Phi-4\nabla_{M}\Phi\nabla^{M}\Phi & =0,\nonumber \\
\frac{1}{\sqrt{-g}}\partial_{M}\left[\sqrt{-g}F^{MNPQ}\right] & =0.\tag{A-3}\label{eq:A-3}
\end{align}
Note that the solution given in (\ref{eq:7}) exactly satisfies the
above equations of motion. The Romand-Romand strength $F_{4}$ needs
to be supported by $N_{c}$ units of flux on $S^{4}$ which satisfies,

\[
\frac{k_{10}N_{c}}{\sqrt{\pi}l_{s}}=\int_{S^{4}}F_{4}.\tag{A-4}
\]
The factor $N_{c}$ arises because in the WSS model we consider a
stack of $N_{c}$ D4-branes.

\section*{Appendix B: The action of the D-brane with non-Abelian excitation}

The action of D-branes with non-Abelian excitations describes the
dynamics of coincident $N$ D$p$-branes. It could be obtained by
T-duality which is the standard technique in string theory. Let us
consider a stack of D$p$-branes in $D$ dimensional spacetime parametrized
by $\left\{ X^{\mu}\right\} ,\mu=0,1...D-1$. Then we use the indices
$a,b=0,1...p$ and $i,j,k=p+1...D-1$ to denote respectively the directions
parallel and vertical to the D$p$-branes in the spacetime. The complete
bosonic action of such D$p$-branes is therefore collected as,

\[
S_{\mathrm{D}_{p}-\mathrm{branes}}=S_{\mathrm{DBI}}+S_{\mathrm{CS}},\tag{B-1}
\]
where

\begin{align}
S_{\mathrm{DBI}}= & -T_{p}\mathrm{STr}\int d^{p+1}\xi e^{-\Phi}\sqrt{-\det\left\{ \left[E_{ab}+E_{ai}\left(Q^{-1}-\delta\right)^{ij}E_{jb}+2\pi\alpha^{\prime}F_{ab}\right]Q_{\ j}^{i}\right\} },\nonumber \\
S_{\mathrm{CS}}= & \mu_{p}\sum_{n=0,1}\int_{\mathrm{D}_{p}\mathrm{-branes}}C_{p-2n+1}\wedge\frac{\left(B+2\pi\alpha^{\prime}F\right)^{n}}{n!},\nonumber \\
Q_{\ j}^{i}= & \delta^{i}\ j+2\pi\alpha^{\prime}\left[\varphi^{i},\varphi^{k}\right]E_{kj},\ E_{\mu\nu}=g_{\mu\nu}+B_{\mu\nu}.\tag{B-2}\label{eq:B-2}
\end{align}
Here $g_{\mu\nu},B_{\mu\nu}$ is the metric of the $D$ dimensional
spacetime and the 2-form field respectively. The ``STr'' refers
to the ``symmetric trace'' and $F$ is the non-Abelian gauge field
strength defined as $F_{ab}=\partial_{[a}A_{b]}+A_{[a}A_{b]}$. The
transverse modes of the D$p$-branes are denoted as $\varphi^{i}$
's which are in fact the ``T-dualitized'' coordinates given by $2\pi\alpha^{\prime}\varphi^{i}=X^{i}$.
Note that we have chosen the ``static gauge'' throughout the manuscript
in order to gauge away the 2-form field $B$ i.e. $B_{\mu\nu}=0$
and defined $\varphi^{i=9}\equiv\Psi$ in (\ref{eq:32}) since there
is only one transverse coordinate for the case of $\mathrm{D}8/\overline{\mathrm{D}8}$-branes.

By keeping these in mind, the DBI action in (\ref{eq:B-2}) could
be expanded as,

\begin{equation}
S_{\mathrm{BDI}}=-T_{p}\mathrm{Tr}\int d^{p+1}\xi e^{-\Phi}\sqrt{-g}\left[1+\frac{1}{4}\left(2\pi\alpha^{\prime}\right)^{2}F_{ab}F^{ab}+\frac{1}{2}D_{a}\varphi^{i}D_{a}\varphi^{i}+\frac{1}{4}\left[\varphi^{i},\varphi^{j}\right]^{2}\right]+\mathrm{high\ orders}.\tag{B-3}
\end{equation}
Note the gauge field $A_{a}$ and scalar field $\varphi^{i}$ 's are
all in the adjoint representation of $U\left(N\right)$.

As a D-brane is usually supersymmetric, there may be superpartner
terms to the actions given in (\ref{eq:B-2}). Nonetheless we do not
attempt to give the explicit action of those fermionic superpartners
since they are irrelevant to the calculations in this manuscript.
However in the WSS model, one might find it remains to have supersymmetry
on the $N_{f}$ flavour branes below the scale of $M_{KK}$ although
the $N_{c}$ D4-branes are compactified (non-supersymmetric) \cite{key-42}.
Accordingly it forces us to include the superpartners of the mesons
on the flavour branes. This becomes indeed a theoretical issue of
this model because the meson spectrum containing superpartners would
be unrealistic. In order to figure out this problem in this model,
we are motivated by the mechanisms in the compactification of the
$N_{c}$ D4-branes and the processes are as follows. First let us
compact the time direction in 10d solution (\ref{eq:7}) on a circle
with the period/mass scale $\delta t\sim M_{T}^{-1}\gg1$. Then we
require the boundary conditions of the bosonic and the supersymmetrically
fermionic fields on D8-brane are periodic and anti-periodic respectively.
So the supersymmetric fermions become massive thus decoupled below
the scale $M_{T}$. Since the mass scale $M_{T}$ corresponds to the
temperature in the dual field theory, it means there is not any superpartners
of the mesons below the (very low) temperature $T=M_{T}\ll1$. Hence
the meson spectrum close to zero temperature is non-supersymmetric
unless we take the limitation of $\delta t\rightarrow\infty$ i.e.
the case of  zero-temperature. This scheme provides a non-suppersymmetric
meson spectrum at very low but finite temperature and it would be
reasonable since zero-temperature case in realistic physics is out
of reach thus it could be supersymmetric.

\section*{Appendix C: The interactions of glueball and $\pi,\rho$ mesons}

\subsection*{C1. The exotic scalar glueball}

Let identify the lowest modes in (\ref{eq:24}) as vector $\rho$
meson and scalar $\pi$ meson in the sector of the light flavor, then
substitute (\ref{eq:42}) (\ref{eq:43}) into the Yang-Mills action
in (\ref{eq:21}) and perform the integration over $Z$, it leads
to the following 4d effective action for the interaction of $\pi,\rho$
meson and the exotic scalar glueball,

\begin{align*}
S_{G_{E}}^{\pi-\rho}= & -\mathrm{Tr}\int d^{4}x\bigg\{ c_{1}\left[\frac{1}{2}\partial_{\mu}\pi\partial_{\nu}\pi\frac{\partial^{\mu}\partial^{\nu}}{M_{E}^{2}}G_{E}+\frac{1}{4}\left(\partial_{\mu}\pi\right)^{2}\left(1-\frac{\partial^{2}}{M_{E}^{2}}\right)G_{E}\right]\\
 & +c_{2}M_{KK}^{2}\left[\frac{1}{2}\rho_{\mu}\rho_{\nu}\frac{\partial^{\mu}\partial^{\nu}}{M_{E}^{2}}G_{E}+\frac{1}{4}\left(\rho_{\mu}\right)^{2}\left(1-\frac{\partial^{2}}{M_{E}^{2}}\right)G_{E}\right]\\
 & +c_{3}\left[\frac{1}{2}\eta^{\sigma\lambda}\partial_{[\mu}\rho_{\sigma]}\partial_{[\nu}\rho_{\lambda]}\frac{\partial^{\mu}\partial^{\nu}}{M_{E}^{2}}G_{E}-\frac{1}{8}\partial_{[\mu}\rho_{\nu]}\partial^{[\mu}\rho^{\nu]}\left(1+\frac{\partial^{2}}{M_{E}^{2}}\right)G_{E}\right]+c_{4}\frac{3}{2M_{E}^{2}}\rho_{\mu}\partial^{[\mu}\rho^{\nu]}\partial_{\nu}G_{E}\\
 & +c_{5}\left[\partial_{\mu}\pi\left[\pi,\rho_{\nu}\right]\frac{\partial^{\mu}\partial^{\nu}}{M_{E}^{2}}G_{E}+\frac{1}{2}\partial_{\mu}\pi\left[\pi,\rho^{\mu}\right]\left(1-\frac{\partial^{2}}{M_{E}^{2}}\right)G_{E}\right]\\
 & +\left[\frac{1}{2}\tilde{c}_{1}\left(\partial_{\mu}\pi\right)^{2}+\frac{1}{2}\tilde{c}_{2}M_{KK}^{2}\left(\rho_{\mu}\right)^{2}+\frac{1}{4}\tilde{c}_{3}\partial_{[\mu}\rho_{\nu]}\partial^{[\mu}\rho^{\nu]}+\tilde{c}_{5}\partial_{\mu}\pi\left[\pi,\rho^{\mu}\right]\right]G_{E}\bigg\},\tag{C-1}
\end{align*}
where the coefficients $c_{i}$'s and $\tilde{c}_{i}$'s are,

\begin{align*}
c_{1} & =\int dZ\frac{\bar{H}_{E}}{\pi K},\ \ \ c_{2}=2\kappa\int dZK\left(\psi_{1}^{\prime}\right)\bar{H}_{E},\ \ \ c_{3}=2\kappa\int dZK^{-1/3}\left(\psi_{1}\right)^{2}\bar{H}_{E},\\
c_{4} & =2\kappa M_{KK}^{2}\int dZ\frac{20ZK}{\left(5K-2\right)^{2}}\psi_{1}\psi_{1}^{\prime}H_{E},\ \ \ ,c_{5}=\int dZ\frac{\psi_{1}\bar{H}_{E}}{\pi K},\ \ \ \tilde{c}_{1}=\int dZ\frac{H_{E}}{4\pi K},\\
\tilde{c}_{2} & =\frac{1}{2}\kappa\int dZK\left(\psi_{1}^{\prime}\right)^{2}H_{E},\ \ \ \tilde{c}_{3}=\frac{1}{2}\kappa\int dZK^{-1/3}\left(\psi_{1}\right)^{2}H_{E},\ \ \ \tilde{c}_{5}=\int dZ\frac{\psi_{1}H_{E}}{4\pi K}.\tag{C-2}
\end{align*}

\subsection*{C2. The dilatonic and tensor glueball}

The action for the interaction of dilatonic glueball and $\pi,\rho$
mesons could be obtained by plugging the (\ref{eq:14}) into action
(\ref{eq:21}) then performing the integration over $Z$. We collect
the resultantly quadratic terms in $\pi,\rho$ as,

\begin{align*}
S_{G_{D}}^{\pi-\rho}= & \mathrm{Tr}\int d^{4}x\bigg\{\frac{1}{2}d_{1}\partial_{\mu}\pi\partial_{\nu}\pi\left(\eta^{\mu\nu}-\frac{\partial^{\mu}\partial^{\nu}}{M_{D}^{2}}\right)G_{D}+\frac{1}{2}d_{2}M_{KK}^{2}\rho_{\mu}\rho_{\nu}\left(\eta^{\mu\nu}-\frac{\partial^{\mu}\partial^{\nu}}{M_{D}^{2}}\right)G_{D}\\
 & +\frac{1}{2}d_{3}\eta^{\lambda\sigma}\partial_{[\mu}\rho_{\sigma]}\partial_{[\nu}\rho_{\lambda]}\left(\eta^{\mu\nu}-\frac{\partial^{\mu}\partial^{\nu}}{M_{D}^{2}}\right)G_{D}+d_{5}\partial_{\mu}\pi\left[\pi,\rho_{\nu}\right]\left(\eta^{\mu\nu}-\frac{\partial^{\mu}\partial^{\nu}}{M_{D}^{2}}\right)G_{D}\bigg\},\tag{C-3}
\end{align*}
where the coefficients are,

\[
d_{1}=\int dZ\frac{H_{D}}{\pi K},d_{2}=2\kappa\int dZK\left(\psi_{1}^{\prime}\right)^{2}H_{D},d_{3}=2\kappa\int dZK^{-1/3}\left(\psi_{1}\right)^{2}H_{D},d_{5}=\int dZ\frac{\psi_{1}H_{D}}{\pi K}.\tag{C-4}
\]
And similarly, the interaction action of tensor glueball and $\pi,\rho$
mesons is given as,

\begin{align*}
S_{G_{T}}^{\pi-\rho}= & \mathrm{Tr}\int d^{4}x\bigg\{\frac{1}{2}t_{1}T^{\mu\nu}\partial_{\mu}\pi\partial_{\nu}\pi+\frac{1}{2}t_{2}M_{KK}^{2}T^{\mu\nu}\rho_{\mu}\rho_{\nu}+\frac{1}{2}t_{3}T^{\mu\nu}\eta^{\lambda\sigma}\partial_{[\mu}\rho_{\sigma]}\partial_{[\nu}\rho_{\lambda]}\\
 & +t_{5}\partial_{\mu}\pi\left[\pi,\rho_{\nu}\right]T^{\mu\nu}\bigg\},\tag{C-5}
\end{align*}
with the coefficients $t_{i}=\sqrt{6}d_{i}$.

\end{document}